\documentclass[reprint,amsmath,amssymb,aps,nofootinbib,superscriptaddress]{revtex4-1}
\usepackage{booktabs}

\usepackage[pdftex,colorlinks]{hyperref}
\usepackage[dvipsnames]{xcolor}
\definecolor{phthaloblue}{rgb}{0.0, 0.06, 0.54}

\hypersetup{
    colorlinks=true,
    linkcolor=magenta,
    citecolor=blue,
    filecolor=MidnightBlue,
    urlcolor=blue,
    }
\usepackage[pdftex]{graphicx}
\usepackage{bm}
\usepackage{eucal}  
\usepackage{mathrsfs}
\usepackage{cancel}
\usepackage{here}
\usepackage{comment,braket}
\usepackage{epsf}
\usepackage{xcolor} 
\usepackage{amsmath}
\usepackage{graphics}
\usepackage{amsfonts}
\usepackage{amssymb}
\usepackage{latexsym}
\usepackage{color}
\usepackage{natbib}
\usepackage[normalem]{ulem}
\input{colordvi.tex}
\usepackage{feynmp}
\usepackage[utf8]{inputenc}
\DeclareUnicodeCharacter{2212}{-}
\usepackage{orcidlink}
\usepackage{multirow}
\usepackage{subcaption}
\usepackage{enumitem}
\newcommand{\beq}{\begin{equation}}
\newcommand{\eeq}{\end{equation}}
\usepackage{bm}

\newcounter{notecountRJ}[section]

\def\lsim{\mathrel{\raise.3ex\hbox{$<$\kern-.75em\lower1ex\hbox{$\sim$}}}}
\def\gsim{\mathrel{\raise.3ex\hbox{$>$\kern-.75em\lower1ex\hbox{$\sim$}}}}

\usepackage{booktabs}
\usepackage{orcidlink}

\begin{document}

\title{Exploring Ultralight Dark Matter Self-Coupling via the Gravitational Wave Background}

 \author{Pratick Sarkar \orcidlink{0009-0000-8160-0734}}
 \email{spsps2523@iacs.res.in}
 \affiliation{School of Physical Sciences, Indian Association for the Cultivation of Science,\\ 2A \& 2B Raja S.C. Mullick Road, Kolkata-700032, India}

\begin{abstract}

Supermassive black hole (SMBH) binary mergers are a primary source of the
stochastic gravitational wave background (SGWB) in the nanohertz band, now
being actively probed by pulsar timing arrays (PTAs).
We investigate how ultralight dark matter (ULDM) with a quartic
self-interaction, forming solitonic cores around SMBHs, modifies the SGWB
through dynamical friction on the inspiralling binary.
Solving the Gross--Pitaevskii--Poisson (GPP) equations numerically for halo
masses $M_{\rm halo}\sim 10^{13}\,M_{\odot}$, we compute self-consistent
solitonic density profiles across a range of dimensionless self-coupling
parameters $\hat{\lambda}$, covering both attractive and repulsive
self-interactions.
We find that soliton-induced dynamical friction imprints a characteristic
suppression feature in the GW strain spectrum, whose frequency location and
depth are sensitive to both the ULDM mass $m$ and the coupling $\hat\lambda$.
For SMBH masses $M_{\bullet}\sim 10^{8.5}\,M_{\odot}$ and ULDM masses in the
range $m\sim(3\times10^{-22}-2\times10^{-21})\,\mathrm{eV}$, current PTA
observations from NANOGrav, EPTA, and PPTA may place qualitative limits on the attractive and repulsive 
self-coupling of ULDM particles.
ULDM masses exceeding ${\sim}\,1.6\times10^{-21}\,\mathrm{eV}$ are
independently excluded by soliton accretion-timescale constraints.
These results demonstrate that the nanohertz GW spectrum provides a 
complementary probe of ULDM self-interactions, operating in a regime 
distinct from conventional particle-scattering or rotation-curve 
analyses.

\end{abstract}

\maketitle

\flushbottom

\section{Introduction}
Compelling experimental evidence, such as galaxy rotation curves, gravitational lensing, and the cosmic microwave background, strongly supports the existence of dark matter (DM)\cite{Rubin:1980zd,Corbelli:1999af,Refregier:2003ct,Allen:2011zs,Rubakov:2019lyf}.
There are numerous reasons to believe that the universe is structured with a non-relativistic, non-baryonic component that does not emit light\cite{Planck:2018vyg}.
Understanding the nature of this component is one of the major challenges in cosmology and astrophysics\cite{Profumo:2019ujg,Bertone:2016nfn}. Extensive efforts have been made, including direct and indirect searches\cite{Graham:2015ouw,Schumann:2019eaa,Gaskins:2016cha,Cirelli:2024ssz} as well as collider studies\cite{Boveia:2018yeb}, to investigate DM.
One prevailing theory posits that DM consists of collision-less, non-relativistic bosons or fermions, collectively referred to as Cold Dark Matter (CDM)\cite{Planck:2018vyg}. This model works well on large scales (on the order of hundreds of Mpc).  However, on smaller scales, some observations deviate from this picture.
For example, the core-cusp problem arises — where DM-only simulations predict dense, cuspy halos at galactic centers in contrast with the observed cored density profiles\cite{Flores:1994gz,Moore:1994yx,Bullock:2017xww,Zavala:2019gpq}. The absence of definitive results from DM detection experiments, combined with these small-scale discrepancies, has motivated the exploration of alternative models beyond the conventional CDM paradigm.
These alternative models aim to preserve CDM’s success on large scales while resolving its shortcomings at smaller scales.

In addition to the standard CDM paradigm, several alternative dark matter frameworks have been proposed to address these challenges, including warm dark matter (WDM)\cite{Bode:2000gq,Viel:2005qj}, self-interacting dark matter (SIDM)\cite{Spergel:1999mh,Tulin:2012wi,Tulin:2017ara}, and fuzzy ultralight dark matter (ULDM). WDM models, characterized by particles with masses in the keV range, can suppress structure formation on small scales but often struggle to simultaneously reproduce both dwarf galaxy cores and Lyman-$\alpha$ forest constraints. SIDM models, on the other hand, invoke elastic scattering between dark matter particles, quantified by the cross-section-to-mass ratio $(\sigma/m)$, which can alleviate central density cusps but may lead to core collapse in dense environments. Among these alternatives, the fuzzy or ultralight dark matter (ULDM) model has attracted particular attention for its ability to introduce quantum mechanical effects at astrophysical scales through the particle’s large de Broglie wavelength\cite{Hu:2000ke,DelPopolo:2016emo,Delgado:2022vnt,Budker:2023sex,Chowdhury:2023xvy,Dave:2023wjq},
\begin{equation}
\lambda_{\text{dB}} = \frac{h}{m v},
\end{equation}
where $m$ is the particle mass and $v$ its characteristic velocity.
For $m \sim 10^{-22}$ eV and $v \sim 10^{-3}c$, $\lambda_{\text{dB}}$ can reach kiloparsec scales
and such particles could occupy a large volume due to their long de Broglie wavelength, exhibiting wavelike behaviour at galactic scales\cite{Marsh:2015wka}. 

Axion-like particles 
(ALPs), arising naturally as pseudo-Nambu--Goldstone bosons in 
string theory compactifications and other beyond-Standard-Model 
frameworks, are among the best-motivated candidates for ULDM in 
this mass range~\cite{Marsh:2015wka,	Hui:2016ltb}. Their extremely small 
masses are technically natural owing to the approximate shift 
symmetry of the underlying potential, making them compelling 
realizations of the fuzzy dark matter paradigm at galactic 
scales.
This wavelike nature of ULDM particles introduces quantum pressure effects that can counteract gravitational collapse, thereby preventing the formation of cuspy density profiles\cite{Hu:2000ke,Garcia:2023abs,Brax:2019fzb,Brax:2019npi}.
At the centers of dark matter halos, ULDM particles can give rise to stable and dense configurations—referred to as soliton cores\cite{Hui:2016ltb,Bar:2019pnz}—which are, however, less dense than the cuspy profiles typically expected in standard CDM scenarios. Thus, ULDM provides a natural explanation for the formation of cored density profiles while remaining consistent with large-scale structure observations.

Since the centers of galaxies harbor massive black holes, their gravitational influence can further affect soliton core formation and stability.
The ULDM that forms this soliton core is described by a scalar field. 
The problem of a scalar field having self-interaction, minimally coupled to gravity in the presence of a black hole, is described by the coupled Gross-Pitaevskii-Poisson equations and has been studied extensively in the literature\cite{GROSS195857,Gross61,Pitaeveskii,2016pqf..book.....B}.
These coupled equations are based on a weak-field, slowly varying, non-relativistic canonical scalar field model for ULDM particles and provide the foundation for modeling solitonic dark matter halos in the presence of black holes.

Numerical simulations show that ULDM halos form solitonic cores whose properties are sensitive to self-interactions \cite{Stallovits:2024cpg}. Studies incorporating a repulsive quartic self-interaction indicate larger, less dense cores and long-lived oscillatory behaviour compared to the non-interacting case \cite{Stallovits:2024cpg,Indjin:2025lqs}. Rotation-curve analyses of dark-matter dominated galaxies suggest that a weak self-interaction can reproduce observed profiles while constraining both the particle mass and coupling strength \cite{Stallovits:2024cpg}. However, observations of ultra-faint dwarf galaxies place tight limits on the particle mass,
required to remain consistent with stellar kinematics and Lyman-$\alpha$ bounds \cite{Teodori:2025rul,Benito:2025xuh}. On cluster scales, gravitational lensing offers complementary tests, with self-interacting dark matter models able to reproduce certain strong-lensing features but being constrained by limits on excessive core collapse \cite{Yang:2021kdf,Gilman:2022ida,GalazoGarcia:2025xjh}. Overall, self-interacting ULDM remains viable within narrow parameter ranges, with ongoing lensing and galactic dynamics studies continuing to refine these constraints.

Recent studies have therefore focused on ULDM  with self-interactions, as explored in works such as \cite{Glennon:2023gfm,Alonso-Alvarez:2024gdz,Blas:2024duy,Teodori:2025rul}. In such models, determining observational constraints on the mass($m$) and self-coupling($\lambda$) of ULDM that forms the soliton core is critically important. In particular, exploring how ULDM solitons interact with compact objects such as supermassive black holes (SMBH) opens a new observational window—especially through gravitational wave signals.

The detection of the stochastic gravitational wave background (SGWB) through pulsar timing array (PTA) observations can provide evidence of SMBH binary mergers.
Currently, several major PTA collaborations are actively searching for SGWB signals. These include the European Pulsar Timing Array (EPTA)\cite{EPTA:2023fyk,EPTA:2023xxk}, Parkes Pulsar Timing Array (PPTA)
\cite{Reardon:2023gzh,Zic:2023gta,Reardon:2023zen}, North American Nanohertz Observatory for Gravitational Waves (NANOGrav)\cite{NANOGrav:2023hfp,NANOGrav:2023hvm}, the Indian Pulsar Timing Array (InPTA)\cite{EPTA:2023fyk}, the Chinese Pulsar Timing Array (CPTA)\cite{Xu:2023wog}, and the International Pulsar Timing Array (IPTA)\cite{InternationalPulsarTimingArray:2023mzf}, which combines data from multiple PTA experiments. PTAs rely on precise timing measurements of highly stable millisecond pulsars to detect variations in the propagation of light caused by gravitational waves (GWs).
 
The SGWB observed by PTAs aligns well with expectations from inspiraling SMBH binaries\cite{Begelman:1980vb,2013ApJ...764..184M}. These SMBHs, with masses ranging from $10^{5}M_{\odot}$ to $10^{10}M_{\odot}$, are typically found at the centers of galaxies. During hierarchical structure formation, galaxy mergers may lead to the formation of SMBH binaries. In the inspiral stages of their evolution, these binaries emit gravitational waves in the nanohertz frequency band, which fall within the sensitivity range of current PTA observations.
If binary evolution is driven solely by the loss of orbital energy through 
GW emission, the SGWB can be described by a characteristic strain $h_c$ with a power-law dependence
on frequency, $h_{c}(f)\propto f^{-2/3}$\cite{Phinney2001APT}.
However, realistic modeling of the astrophysical environments surrounding the binaries suggests modifications to this power-law\cite{Hu:2023oiu,NANOGrav:2024nmo}. 
Beyond purely astrophysical effects, deviations in the power-law index of the SGWB, driven by new physics scenarios near SMBH binaries, can be explored. In particular, the presence of a dense ULDM environment, such as soliton cores, induces friction on black hole binary, influencing the generation of the SGWB.
During a merger event, when the separation between the SMBHs becomes comparable to soliton radius, both black holes experience a wind of ULDM particles.
This phenomenon, known as dynamical friction\cite{Chandrasekhar:1943ys}, introduces a novel energy loss mechanism that can compete with gravitational wave emission in the nanohertz frequency range.
Consequently, the presence of soliton cores alters the SGWB spectrum. Current PTA data can provide constraints on the modified power-law index of the SGWB, placing limits on the ULDM mass($m$), self -coupling($\lambda$) and soliton mass. 

Several recent works have explored related but distinct aspects of ULDM and
gravitational waves.
Aghaie~et~al.~\cite{Aghaie:2023lan} and Ghoshal~\&~Strumia~\cite{Ghoshal:2023fhh}
constrained ULDM masses using NANOGrav data, but did not incorporate
self-interactions in the soliton profile.
Alonso-\'{A}lvarez~et~al.~\cite{Alonso-Alvarez:2024gdz} demonstrated that
self-interacting dark matter can assist SMBH binary hardening to address the
final parsec problem, but worked within a particle dark matter framework
rather than the wavelike, soliton-forming regime of ULDM.
Glennon~et~al.~\cite{Glennon:2023gfm} and Koo~et~al.~\cite{koo2024:phylett} studied dynamical friction in
self-interacting ULDM but did not connect their results to the observable
SGWB spectrum or derive constraints from PTA data.
The present work fills this gap by:
\begin{enumerate}[label=(\roman*)]
    \item numerically solving the GPP equations with a quartic
          self-interaction term to obtain physically motivated solitonic
          density profiles for a range of both attractive and repulsive
          couplings $\hat\lambda\in[-1,+1]$;
    \item computing the resulting modification to the SGWB strain spectrum
          through soliton-induced dynamical friction on inspiralling SMBH
          binaries; and
    \item mapping the observable suppression features in the nanohertz band
          in the $(m,\,\lambda)$ parameter space of the ULDM particles
          using current PTA observations from NANOGrav, EPTA, and PPTA.
\end{enumerate}
Unlike approaches based on the particle scattering cross-section ratio
$\sigma/m$, our framework begins from a fundamental scalar field action with
a $\phi^{4}$ self-interaction, providing a self-consistent connection between
the microphysical coupling $\lambda$ and the macroscopic solitonic dark matter
density profile. Throughout this work, we assume that the ULDM particles carry no 
charge under Standard Model gauge groups and interact with ordinary 
matter exclusively through gravity.
This makes the nanohertz SGWB a uniquely sensitive probe of ULDM
self-coupling in a regime that is complementary to both galactic dynamics and
gravitational lensing constraints.

The structure of the paper is as follows: Section~\ref{ULDM_Solitons} introduces solitons, formed from ULDM particles around SMBHs. This section discusses the fundamental approach to solving the Gross-Pitaevskii-Poisson equation and deriving solitonic DM profiles. The impact of solitonic DM on the gravitational wave spectrum is explored in Section~\ref{GW_from_merger}. The results are presented in Section~\ref{results}, and the conclusions are drawn in Section~\ref{Discussion}.

Throughout the paper, we frequently use natural units ($c=\hbar=1$) but retain  $c$ and $\hbar$
explicitly in some of the expressions. The reduced Planck mass is denoted by $M_{pl}$ i.e. $M_{pl} =\sqrt{c\hbar/8\pi G_N} \approx 2.4\times 10^{18}\rm GeV$. The symbol $M_{\odot}$  denotes the solar mass, which is $M_{\odot} \approx 2\times 10^{30}$ kg.

\section{ULDM Soliton around central SMBH}\label{ULDM_Solitons}

The literature study of ULDM suggests the existence of a dense, macroscopic core at the center of galaxies, commonly known as a soliton\cite{1993ApJ...416L..71W,Coles:2002sj,Marsh:2015wka,Davies:2019wgi,Bar:2018acw}. This form of dark matter, characterized by an extremely high occupation number, behaves as a classical scalar field and is governed by the wave equation\cite{Tremaine:1979we,Khlopov:1985fch,Hui:2021tkt}. The  classical field theory action for this scalar field $\phi$, is given by

\begin{equation}
    S = \int d^{4}x \sqrt{-g}\left(\frac{M_{pl}^2}{2} R - \frac{1}{2}g^{\mu\nu}\partial_{\mu}\phi \partial_{\nu}\phi -U(\phi)\right)
\end{equation}

where the potential has the form
   \begin{equation}
    U(\phi)=\frac{m^2\phi^2}{2}+\frac{\lambda\phi^4}{4!}
\end{equation} 

The coefficient $\lambda$ here denotes the self-coupling parameter of the ULDM particles.

By varying the action with respect to $\phi$, we obtain the equation of motion:

\begin{equation}
   \partial_{\alpha}\left(\sqrt{-g}\, g^{\alpha\beta}\,\partial_{\beta}\phi\right) =\sqrt{-g}\, U'(\phi)
\end{equation}
where $U'(\phi)$ denotes the derivative of the potential with respect to $\phi$. 
R is the Ricci scalar, and g denotes $det(g_{\mu\nu})$.

Strong evidence suggests that all large galaxies, and even dwarf galaxies, harbor a central SMBH~\cite{Kormendy:1995er,Kormendy:2013dxa,Reines:2022ste}. In such systems, the SMBH provides the dominant central gravitational potential, within which the ultralight dark matter field can form a gravitationally bound macroscopic soliton~\cite{Ruffini:1971bza,Israel:1967wq,Israel:1967za,Carter:1971zc,Bekenstein:1972ny}. The system of an SMBH embedded within a scalar-field dark matter halo is often described using a spherically symmetric metric~\cite{Cardoso:2022nzc,Figueiredo:2023gas,Zhong:2023xll,Mitra:2023sny}, with the Schwarzschild metric being the simplest example. 

Alternatively, the cosmological ULDM scalar field can be analyzed by adopting the perturbed FRW metric, as discussed in \cite{1993ApJ...416L..71W,Hui:2016ltb,Fan:2016rda}.
In galactic dynamics, the effects of cosmic expansion are negligible. Therefore, we can set the scale factor to unity and the Hubble parameter to zero, leading to the following metric:

\begin{equation}
    ds^{2}=-(1+2\Phi)(d{x^0})^2+(1-2\Phi)\delta_{ij}dx^{i}\,dx^{j}
\end{equation}

within the weak gravity limit, $\Phi \ll 1$, where the Newtonian potential $\Phi$ dominates the gravitational influence\footnote{Anisotropy effects were neglected in this analysis for simplicity.}.

Under this metric, the equation of motion simplifies to:

\begin{equation}
    \partial_{0}^2 \phi-\nabla^{2}\phi+U'(\phi)=2\Phi(2\partial_{0}^2\phi+U'(\phi))
\end{equation}

The massive scalar field $\phi$, which satisfies the Klein-Gordon equation and is minimally coupled to gravity, can be expressed in the non-relativistic limit as \cite{1993ApJ...416L..71W,Marsh:2015wka},

\begin{equation}
    \phi(t,\vec{x})=\frac{1}{\sqrt{2}m}\left(e^{-imt}\Psi(t,\vec{x})+\text{c.c.}\right)
\end{equation}

where $m$ represents the mass of the scalar field, and $\Psi(t,\vec{x})$ encapsulates its dynamical behaviour. Here, the complex field $\Psi$ varies slowly in both space and time, satisfying the conditions $|\nabla\Psi| \ll m|\Psi|$, $|\dot{\Psi}| \ll m|\Psi|$ and $|\ddot{\Psi}| \ll m|\dot{\Psi}|$ .

In the regime of weak gravity and slowly varying fields, $\Psi$ satisfies the Gross-Pitaevskii-Poisson (GPP) equation\cite{GROSS195857,Gross61,Pitaeveskii,Marsh:2015wka,Davies:2019wgi,Bar:2018acw,Bar:2019pnz,Chakrabarti:2022owq}:
\begin{gather}\label{GPP-Eq-full}
i \frac{\partial\Psi}{\partial t} = -\frac{\nabla^{2}\Psi}{2m} + m \Phi \Psi  +\frac{\lambda}{8m^{3}} |\Psi|^{2}\Psi  \\
\nabla^2 \Phi  =4\pi G_{N} |\Psi|^{2}
\end{gather}
We seek spherically symmetric steady-state solutions, assuming an ansatz for\footnote{$\Psi(t,\vec{x})=\chi(\vec{x}) e^{-i\gamma t}$ and for spherical symmetry $\chi(\vec{x})=\chi(r)$} $\Psi(t,r)=\chi(r) e^{-i\gamma t}$, where the ULDM density is given by $\rho(r)=\chi^{2}(r)$. The solution $\chi(r)$ is real, spherically symmetric, spatially localized, and without nodes. These solutions are known as solitons. The influence of a central black hole is incorporated via the term $m\Phi\Psi=m(V+\Phi_{BH})\Psi$, where $V$ includes the self-potential  and the black hole potential is given by, $\Phi_{BH}= -\frac{G_{N}M_{BH}}{r}$. Here, $M_{BH}$ is the black hole mass, and $G_{N}$ is Newton's constant.
In the absence of self-interactions, the system is usually referred to as the Schr\"odinger-Poisson system.

To solve the coupled equations numerically, we introduce dimensionless variables\cite{Davies:2019wgi}: distance $\hat{r}=\frac{r}{\hbar/ mc}$, wave function $\hat{\chi}=\frac{\sqrt{4\pi G_N}\hbar}{mc^2}\chi$, potential $\hat{V}=V/c^2$, eigen energy $\hat{\gamma}=\gamma/mc^2$. We also define dimensionless parameters for the black hole,
\begin{equation}
\hat{\alpha}=\frac{G_N M_{BH} m}{c\hbar}
\end{equation} 
and for the self-interaction, 
\begin{equation}\label{eq:self_coupling}
    \hat{\lambda}=\frac{\lambda}{8}\left(\frac{m}{M_{pl}}\right)^{-2}=\left(\frac{\lambda}{1.35\times10^{-96}}\right)\left(\frac{m}{10^{-21}eV}\right)^{-2}
\end{equation}
Although the quartic self-coupling, $\lambda$ of ultra-light particles is extremely small, 
it can become dynamically relevant in regions of high phase-space density. 
Even such a tiny interaction may significantly affect the stability 
and structure of ULDM halos, such as axion halos, 
as shown in several works~\cite{Chavanis:2011zi,Chavanis:2011zm,Eby:2014fya,Chavanis:2016dab,Levkov:2016rkk,Eby:2017azn,Helfer:2016ljl,Desjacques:2017fmf}.

Using the dimensionless variables, the GPP equation takes the form
\begin{equation}\label{GPP-Eq-dimensionless}
\begin{aligned}
\hat{\gamma} \hat{\chi} &= \left( -\frac{1}{2}\hat{\nabla}^2 + \hat{V} - \frac{\hat{\alpha}}{\hat{r}} + 2\hat{\lambda} \hat{\chi}^2 \right) \hat{\chi}, \\
\hat{\nabla}^2 \hat{V} &= \hat{\chi}^2.
\end{aligned}
\end{equation}

Here, the parameters $\hat{\alpha}$ and $\hat{\lambda}$ are free and chosen to solve the system. The parameter $\hat{\gamma}$ is determined using a shooting method by applying boundary conditions to find solutions without nodes~\cite{Lora:2011yc}. For numerical work, we impose the normalization~ $\hat{\chi}(\hat{r}=0)\sim 1$ along with the boundary conditions:
$\hat{\chi}(\hat{r}=\infty)= 0, \hat{\chi}'(\hat{r}=0)= 0,\hat{V}(\hat{r}=0)= 0, \hat{V}'(\hat{r}=0)= 0 $.
Although $\hat{\chi}(\hat{r}=0)=0$ may also yield solutions, these are not realistic for modeling physical soliton profiles, since the other dimensionless parameters are $\mathcal{O}(1)$.

\begin{figure}[h!]
    \centering
    \includegraphics[width=0.9\linewidth]{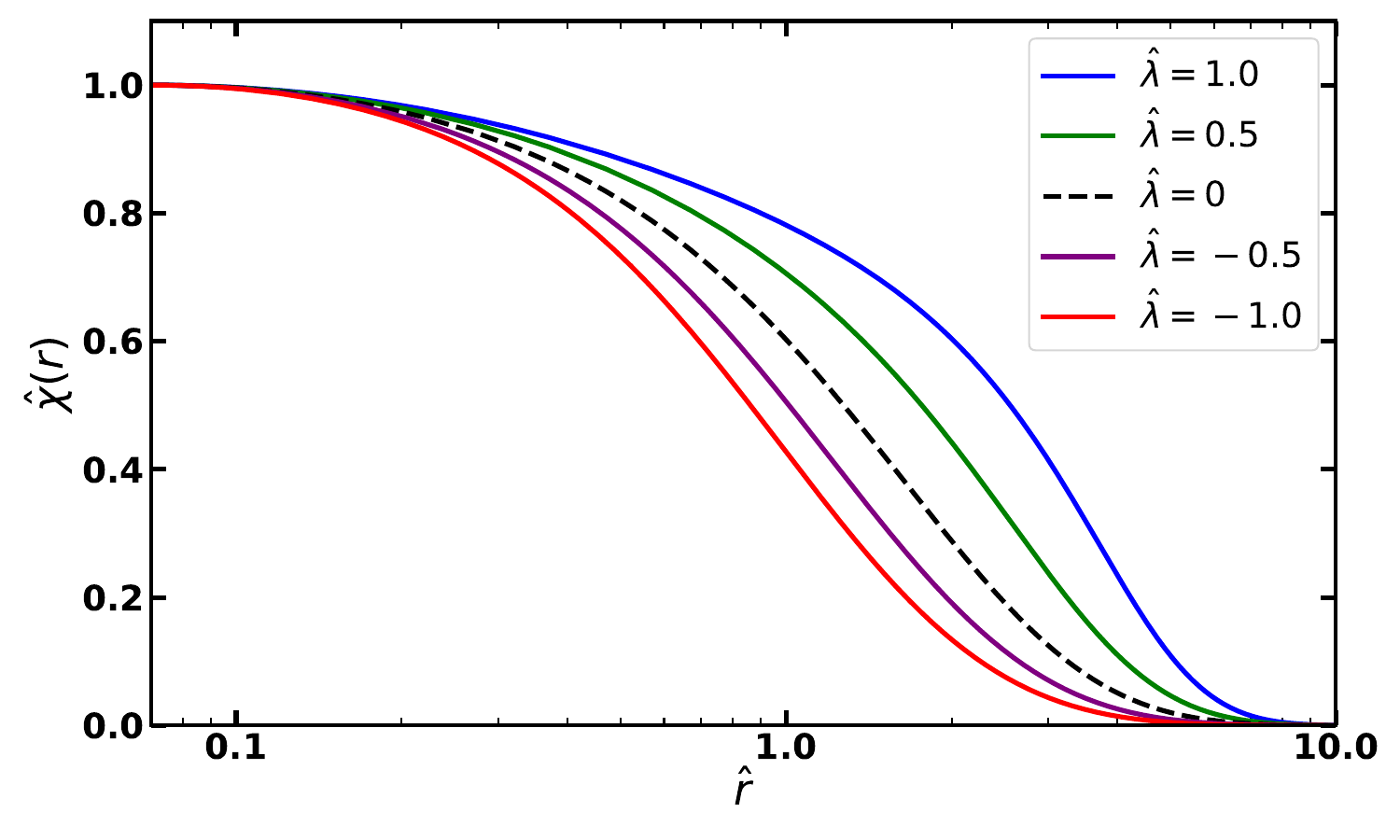}
    \caption{Solutions of the coupled, dimensionless Gross-Pitaevskii-Poisson equation \ref{GPP-Eq-dimensionless} for various values of $\hat{\lambda}$ keeping $\hat{\alpha}$ fixed at 0.48 for a ULDM mass of $10^{-21}$eV.}
    \label{fig:GPP-dimensionless}
\end{figure}

The numerical shooting procedure adopted here follows the standard approach used in solving the dimensionless GPP equations for ultralight dark matter solitons (see, e.g., Refs.~\cite{Marsh:2015wka,Davies:2019wgi}). In particular, the choice of the central normalization $\hat{\chi}(0)=1$ is a numerical convention in the dimensionless formulation, while the physical density profile is recovered after applying the scaling transformation discussed below.

For a fixed SMBH mass parameter $\hat{\alpha}$ corresponding to halo mass $M_{halo}=10^{13}M_{\odot}$(shown in the table~\ref{tab:example_table}), we choose the $\hat{\lambda}$ parameter of $\mathcal{O}(1)$. To solve this system of differential equations, we apply a shooting method. This method integrates the solution from $\hat{r}=0$, aiming to satisfy the asymptotic boundary condition 
$\hat{\chi}(\hat{r}=\infty)=0$. To achieve this, an initial guess for $\hat{\gamma}$ is needed, as only quantized values meet the asymptotic boundary conditions. Using a standard root-finding technique, we determine the smallest  $\hat{\gamma}_{0}$ that satisfies the boundary conditions, giving the 0-node (ground state) solution. Figure~\ref{fig:GPP-dimensionless} illustrates the dimensionless solutions obtained for various values of $\hat{\lambda}$ with a specific $\hat{\alpha}$ value. In the limit of large $\hat{\alpha}$ (representing a large black hole mass), the solution approaches the exponential form characteristic of the hydrogen atom.

A relevant parameter in our analysis is the ratio of the soliton mass to the black hole mass.  
To ensure that the SMBH generates potential in the system,  
this ratio has been kept at \( \lesssim 0.3 \) throughout the study~\cite{Davies:2019wgi},  
allowing us to recover the exponential solutions.

For each unique solution, the total mass of the soliton is given by
\begin{equation}\label{dimensionlessmass}
\begin{split}
    M &= \frac{\hbar c}{G_N m} \int_0^{\infty} \hat{\chi}^2 \hat{r}^2 
        \, d\hat{r} \\
      &= 1.33 \times 10^{11} M_{\odot} 
        \left(\frac{10^{-21} \, \mathrm{eV}}{m}\right) 
        \int_0^{\infty} \hat{\chi}^2 \hat{r}^2 \, d\hat{r}.
\end{split}
\end{equation}
In dimensionless units, the soliton mass is expressed as
\begin{equation}\label{dimension_less_mass}
    \hat{M} \equiv \frac{G_N M m}{\hbar c} = \int_0^{\infty} \hat{\chi}^2 \hat{r}^2 d\hat{r}.
\end{equation}

The central ULDM density is derived from $\chi = \frac{m c^2}{\sqrt{4 \pi G_N} \hbar} \hat{\chi}$ which is given by
\begin{equation}
    \rho  = 2 (M_\text{pl} m)^2\,\hat{\chi}^{2} \sim 10^{16} \left(\frac{m}{10^{-21} \, \text{eV}}\right)^2 \,\hat{\chi}^{2}\,M_{\odot}/\text{pc}^3,
\end{equation}
Here this value is estimated without taking self-interactions. This density corresponds to a very large mass within a volume of $(\text{pc})^3$. Consequently, solutions assuming $\hat{\chi}(\hat{r}=0)\sim 1$ yield excessively high densities.

In the absence of a central SMBH, setting $\hat{\alpha}=0$ in the GPP equation, the Poisson equation implies $\hat{V} \sim \hat{L}^2 \hat{\chi}^2$ for a system of size $\hat{L}$ (in units of $\hbar/mc$). This leads to $\hat{L} \sim \hat{\chi}^{-1/2}$. If $\hat{\chi} \leq 1$, then $\hat{L}$ remains $\mathcal{O}(1)$. In dimensionful units without the central BH the size of the soliton will be of the order of ULDM Compton wavelength, $\hbar/mc(i.e. \frac{\hbar}{mc} \times \hat{L})$.

When a central BH is present ($\hat{\alpha} > 0$), for $\hat{\alpha} \ll 1$, the $\hat{V}\hat{\chi}$ term dominates over $\frac{\hat{\alpha}}{\hat{r}} \hat{\chi}$ in the GPP equation, keeping the soliton size $\mathcal{O}(1)$. When $\frac{\hat{\alpha}}{\hat{r}} \hat{\chi}$ dominates, the size becomes $\hat{L} \sim 1/\hat{\alpha}$, which requires $\hat{\alpha}$ to be at least $\mathcal{O}(1)$. Thus, the dimensionless size $\hat{L}$ remains $\mathcal{O}(1)$ in either case.

The soliton mass, determined from Eq.~\ref{dimension_less_mass}, is $\mathcal{O}(1)$ in dimensionless units, corresponding to a physical mass of $10^{11} M_{\odot}$, which exceeds the typical mass of a soliton in galaxies within the mass range of $M_{halo}\sim(10^{12}-10^{13}) M_{\odot}$. The gravitational radius of the soliton is
\begin{equation}
    R_G = \frac{G_N M}{c^2} = \hat{M} \frac{\hbar}{mc} \sim \frac{\hbar}{mc},
\end{equation}
comparable to the soliton size. Real solitons are expected to be larger, less dense, and much lighter. Mapping these estimated soliton densities and sizes to physical values requires scaling of the solutions and parameters in the GPP equation, which is discussed in the next section.

\subsection{Scaling relation }
The GPP system of equations has the scaling symmetry\cite{Marsh:2015wka,Davies:2019wgi,Chakrabarti:2022owq}
\begin{gather*}
    \hat{\chi} \rightarrow \frac{1}{s^2}\hat{\chi}, \hspace{4mm}
    \hat{V} \rightarrow \frac{1}{s^2}\hat{\Phi},\hspace{4mm}
    \hat{\alpha} \rightarrow  \frac{1}{s}\hat{\alpha} ,\hspace{4mm}
    \hat{\gamma} \rightarrow  \frac{1}{s^2}\hat{\gamma} ,\hspace{4mm}\\
    \hat{r} \rightarrow s\hat{r},\hspace{4mm}
    \hat{\lambda} \rightarrow s^{2} \hat{\lambda},\hspace{4mm}
    \hat{M} \rightarrow  \frac{1}{s}\hat{M} \label{masslambda}
\end{gather*}
s being a scaling parameter, which can be used to transform the dimensionless numerical solutions and parameters to their dimensionful physical values. Each of the term in GPP equation as well as the density scales as $1/s^{4}$. The ratio of the soliton mass and the black hole mass, $M/M_{BH}$ is invariant under the scaling and is the primary parameter which stabilizes the shape of the soliton.

Now we have to model the density profile of the soliton for a given halo mass as a function of ULDM mass using the numerical solutions as discussed in the previous section.

Over cosmic timescales, the stability of the solitonic core at the center of a galaxy depends on whether it can resist being accreted by the central SMBH. For ULDM particles, the timescale for this accretion is given by \cite{Barranco:2011eyw,Barranco:2012qs},
\begin{equation}\label{accr_time}
    t_{accr}=5.6\times 10^{12}\rm yr \Bigg(\frac{M_{BH}}{10^{8}\rm M_{\odot}}\Bigg)^{-5} \Bigg(\frac{m}{10^{-21}}\Bigg)^{-6}
\end{equation}
To ensure the solitonic core remains intact today, this accretion timescale
is expected to be sufficiently long over cosmological durations. 
Since $t_{accr}\propto m^{-6}$, increasing the mass of the ULDM significantly shortens the accretion time. Therefore, requiring that the soliton has not yet been absorbed by the SMBH leads to an upper bound on the ULDM particle mass.

\begin{figure*}[ht!]
    \centering
    \includegraphics[width=0.48\linewidth]{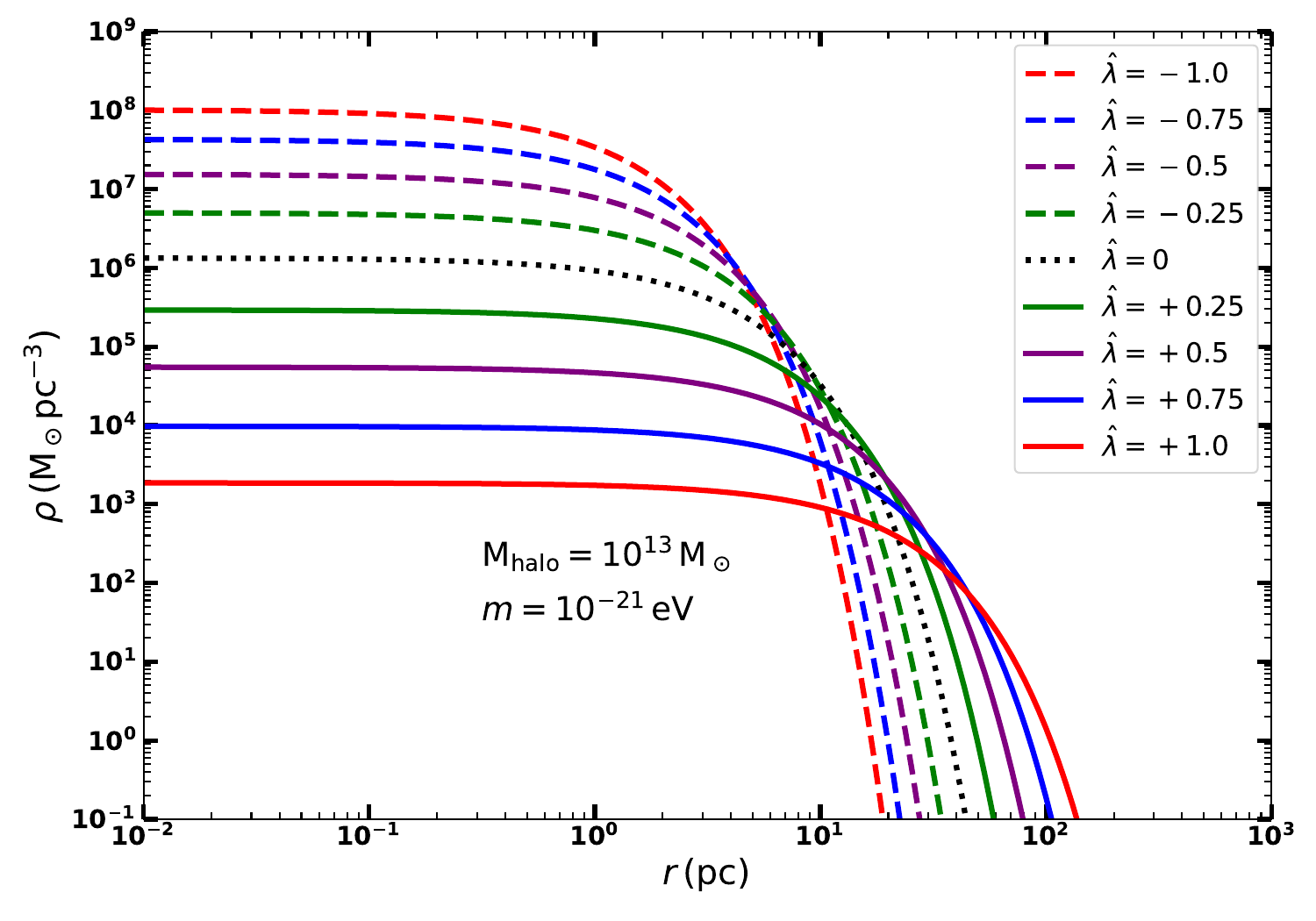}
     \includegraphics[width=0.48\linewidth]{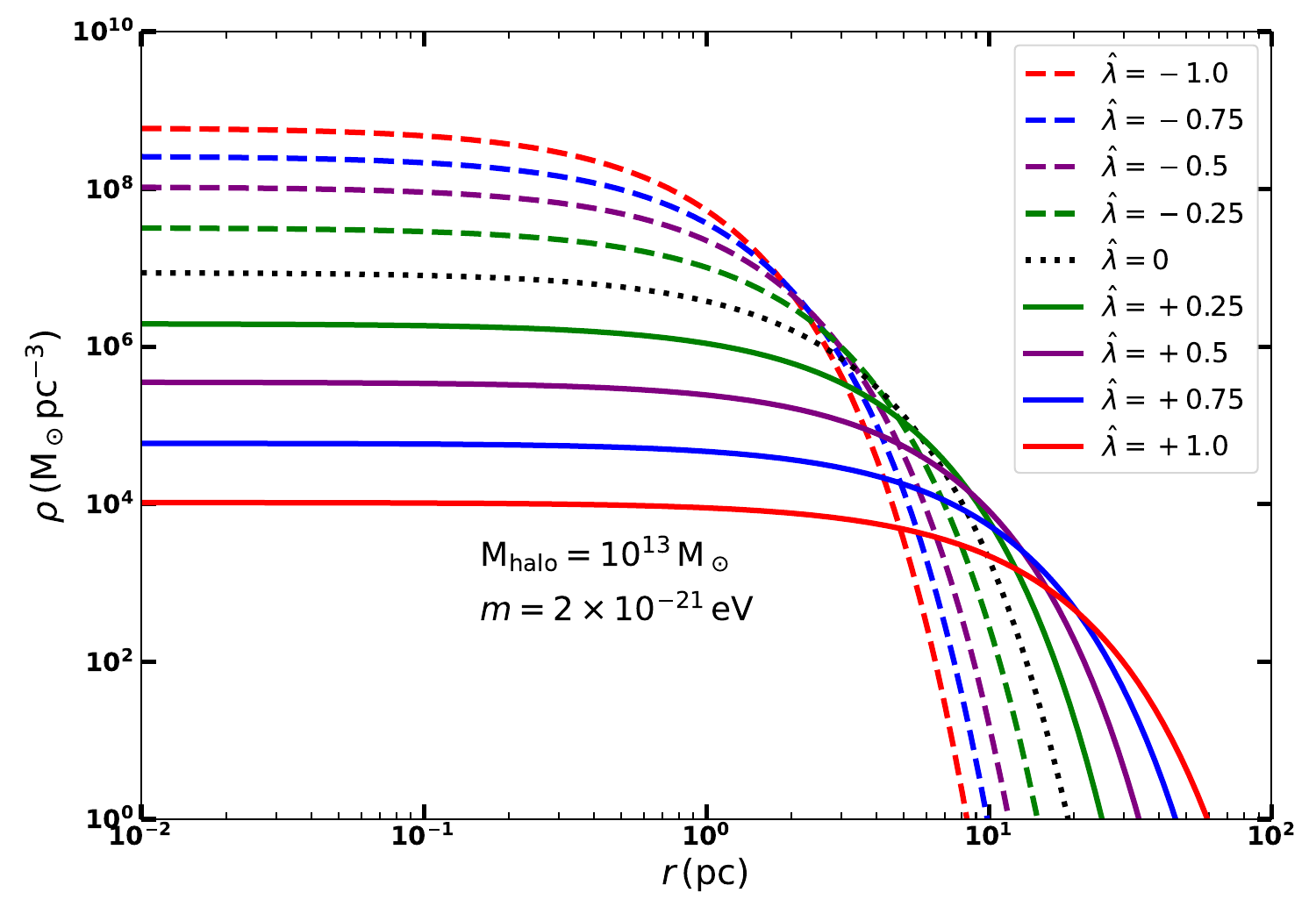}
    \caption{Numerical solutions to the GPP equation are presented for a halo mass of $10^{13}\rm M_{\odot}$ and ULDM masses of $10^{-21}\rm eV$(Left) and $2\times 10^{-21}\rm eV$(Right). The DM density profiles, shown in different colors, illustrate the effect of varying the dimensionless self-coupling parameter $\hat{\lambda}$, each with distinct characteristic signatures. The GPP equation was solved in dimensionful units, incorporating the gravitational influence of the SMBH through the appropriate values of the $\hat{\alpha}$ parameter corresponding to the given halo mass. }
    \label{fig:Density_profile}
\end{figure*}

Once the SMBH completes its growth, it may dominate the central mass of the halo, exerting a force on the system's center. For DM-only simulations, the mass of the soliton $M_{sol}$ is related to the galactic halo mass via the relation\cite{Schive:2014hza}
\begin{equation}\label{sol-halo mass relation}
    M_{sol} = 2.67\times 10^{8} \left(\frac{M_{halo}}{10^{13}M_{\odot}}\right)^{1/3} \left(\frac{m}{10^{-21}eV}\right)^{-1}M_{\odot}
\end{equation}
Nevertheless, this relation was obtained in the absence of SMBHs: the simulation was performed by fixing the ULDM mass at $m=10^{-22}$~eV, with halo masses in the range $M_{halo}=(10^{9}-5\times 10^{11})M_{\odot}$. For $\hat{\lambda}=0$, our treatment reproduces the SMBH-modified equilibrium solutions of~\cite{Davies:2019wgi} through their tabulated values of the dimensionless parameter $\hat{\alpha}$, and we extend this framework by introducing bosonic self-interactions into the GPP equation. To the best of our
knowledge, no numerical calibration currently exists for systems including both
an SMBH and self-interactions; we therefore retain the standard soliton–halo
relation of Eq.~\ref{sol-halo mass relation}  solely as a benchmark normalization between the halo and
soliton masses, and the effects of attractive or repulsive self-interactions are
then determined self-consistently from the modified equilibrium solutions. This
provides a controlled extension of the existing SMBH–soliton framework while preserving the established
SMBH treatment.

We took the parameter 
$\hat{\alpha}$ according to a particular halo mass~$10^{13} M_{\odot}$, relevant to this study (shown in table~\ref{tab:example_table}), that produces a solution consistent with the predicted soliton mass for the given halo mass. Next, we convert the dimensionless density profile into physical units and apply the scaling symmetry to adjust the soliton solution, so that its total mass aligns with the soliton-halo mass relation. This is done by comparing the mass of the scaled numerical solution to the soliton-halo mass relation in Eq.~\ref{sol-halo mass relation} for the given halo mass. The corresponding scaling factor is given by

\begin{equation} 
s = \frac{c\hbar}{G_{N} m M_{\text{sol}}} \hat{M}= \frac{1.33\times 10^{11}}{2.67\times10^{8}}\left(\frac{M_{\text{halo}}}{10^{13}M_{\odot}}\right)^{-1/3} \hat{M}
\end{equation}
Then using this $s$ , we can scale the numerical solutions and other relevant quantities discussed above. For every initial value of $\hat{\lambda}$, we will get the $s$ parameter for chosen $\hat{\alpha}$. The variation of scaling, $s$ with respect to $\hat{\lambda}$ is shown in \ref{Appd_scaling}. For attractive self-interactions, the $s$ parameter becomes larger in comparison to repulsive self-interaction. This is obvious as the attractive self-interaction makes the soliton core more dense and less in size as shown in Fig.~\ref{fig:Density_profile}

Our study is confined to solitons composed of ULDM particles with masses in the range $\sim (5\times 10^{-22}~\text{eV} - 2\times 10^{-21}~\text{eV})$. 
We further restrict our analysis to galaxies with halo masses of $M_{halo}\sim 10^{13}\rm M_{\odot}$, a mass range characteristic of massive early–type galaxies which constitute SMBHs at their centers. 

\subsection{Density Profiles}\label{Density profiles}

In addition to the soliton-halo mass relation, the SMBH mass is also correlated with the halo mass according to \cite{Bandara:2009sd}:
\begin{equation}\label{halo_smbh_mass_relation}
    \log_{10} (M_{BH}/M_{\odot}) = 8.18 + 1.55 [\log_{10}(M_{halo}/M_{\odot})-13.0]
\end{equation}
allowing us to predict the mass of the SMBH and the soliton core for a given halo mass. Assuming the soliton-halo mass relation holds true in the presence of an SMBH,
the gravitational potential of the SMBH modifies the soliton density profile.

We compute the density profiles in the presence of the SMBH, incorporating its influence through the parameter $\alpha$. The coupled GPP equation, with fixed halo mass, $M_{halo}$ and the parameter $\alpha$, is solved numerically using appropriate boundary conditions. These calculations yield density profiles for various values of the $\lambda$ parameter while keeping $\alpha$ constant. From these solutions, the density is obtained by squaring the soliton solution $\chi$. The ULDM density in physical units is then given by
\begin{equation}\label{density_eq}
    \rho = 4.1\times 10^{16} \left(\frac{m}{10^{-21}}\right)^{2} \hat{\chi}^{2} M_{\odot}/pc^{3}
\end{equation}
Finally, we apply the scaling symmetry relations detailed in the \ref{Appd_scaling} to rescale the soliton density so that its total mass matches the value predicted by the soliton-halo mass relation.

Fig.~\ref{fig:Density_profile} displays the density profiles for different values of $\hat{\lambda}$ while keeping the halo mass fixed at $10^{13}\rm M_{\odot}$.The left panel corresponds to a ULDM mass of $10^{-21}$eV while the right panel represents profiles for a ULDM mass of $2\times10^{-21}$eV. As shown in the figure, an attractive self-coupling($\hat{\lambda}<0$) results in more compact solitons—denser and smaller in size (dashed curves). In contrast, a repulsive self-coupling ($\hat{\lambda}>0$) leads to solitons that are less dense and larger in size (solid curves). The black dotted lines represent solitons with no self-coupling between ULDM particles.

The choice of halo mass $M_{\rm halo} = 10^{13}\,M_{\odot}$ is motivated 
by the requirement that the inspiralling SMBH masses fall within the range 
most sensitive to current PTA observations.
The SGWB signal in the nanohertz band receives its dominant contribution 
from SMBH binaries with masses $M_{\bullet} \sim 10^{8.5}\,M_{\odot}$, 
which are hosted in massive early-type galaxies with halo masses 
$M_{\rm halo} \sim 10^{13}\,M_{\odot}$, as dictated by the 
black hole--halo mass relation of Eq.~\ref{halo_smbh_mass_relation}.
Increasing the halo mass beyond this value pushes the central SMBH mass 
to $M_{\bullet} \gtrsim 10^{9}\,M_{\odot}$, which raises the overall 
amplitude of the GW strain spectrum beyond the range constrained by 
current PTA data.
Conversely, reducing the halo mass lowers the SMBH mass and 
consequently suppresses the strain amplitude below the PTA sensitivity 
threshold, rendering the signal undetectable with present observations.
The halo mass $M_{\rm halo} = 10^{13}\,M_{\odot}$ therefore represents 
the optimal choice for probing ULDM self-coupling signatures within the 
nanohertz sensitivity window of NANOGrav, EPTA, and PPTA.

\section{DM influence on Gravitational Wave Background}\label{GW_from_merger}

Thus far, we have characterized solitonic DM profiles composed of ULDM with self-interactions, concentrated at the centers of galaxies. In this section, we study binary mergers of SMBH that occur within such solitonic environments. Our treatment assumes that the solitonic core evolves in the gravitational potential of the central SMBH.
This assumption is valid in scenarios where the soliton mass is subdominant compared to the black hole, specifically when the mass ratio \( M_{\mathrm{sol}} / M_{\mathrm{BH}} \lesssim 0.3 \) \cite{Davies:2019wgi,Bar:2019pnz}. In this regime, the SMBH dominates the gravitational potential.

In our analysis of the SGWB from SMBH inspirals embedded in ULDM solitons, we neglect the effects of tidal disruption or deformation of the solitonic cores. This is justified on several grounds.
Tidal interactions between a solitonic DM core and the binary SMBH are generally suppressed due to the small gravitational coupling parameter, $M_{\bullet} m \sim 10^{-3}$ (where $M_{\bullet}$is SMBH mass in the range $\sim (10^{8}-10^{9}) M_{\odot}$ and $m$ is ULDM mass in the range \( m \sim 10^{-21}\)) introduced in Ref.\cite{Cardoso:2020hca}.

In many astrophysical scenarios, the soliton is not fully self-gravitating  but is instead stabilized by its host SMBH and the soliton behaves as a gravitationally intact bound state susceptible against the strong tidal disruption. The companion SMBH in the binary introduces only a weak tidal field at typical inspiral separations, and the deformation remains subdominant.
Numerical and analytical studies (e.g.,~\cite{Cardoso:2020hca,Baumann:2018vus}) further support that tidal effects are suppressed unless the binary reaches the very late stage of merger. Therefore, throughout most of the inspiral epoch relevant to the nanohertz GW spectrum observed by PTAs, solitonic cores remain gravitationally intact, and tidal disruption can be safely neglected in our present framework. Nonetheless, such effects could become relevant in high-precision studies of late-stage binary evolution and merit further investigation and lie beyond of the scope of present work.

To model the influence of the dark matter environment on the binary evolution, we include the effect of dynamical friction from the ULDM distribution. In particular, the outer regions of the soliton contribute most significantly to the gravitational drag experienced by the SMBH. Before summing over all merger events for a fixed galaxy mass across cosmological distances, we first analyze a binary system consisting of two equal-mass SMBHs embedded within solitonic cores~\cite{Ghoshal:2023fhh,Aghaie:2023lan,Tomaselli:2024ojz}.

We consider a binary system of black holes with masses \( M_1 \) and \( M_2 \), moving in circular orbits about their common center of mass, with a binary separation \( r \). The angular orbital frequency \( \omega \) is given by Kepler's third law,
\begin{equation}
\omega = \sqrt{\frac{G_N (M_1 + M_2)}{r^3}}.
\end{equation}

In the nanohertz frequency band probed by PTAs, SMBH binaries are typically widely separated compared to their Schwarzschild radii. For one of the black holes with mass \( M_1 \), the Schwarzschild radius is given by \( R_{\text{Sch}} = 2G_N M_1 \). At such separations, the binary dynamics remain well within the Newtonian regime. The corresponding orbital velocity is
\begin{equation}
v_{\rm orb} = \omega r \sim \sqrt{\frac{R_{\text{Sch}}}{r}} \ll 1,
\label{vel_kepler}
\end{equation}
assuming equal-mass binaries.

The dominant mechanism for energy loss depends on the binary separation: at smaller separations, GW emission is the primary driver of orbital decay, whereas at larger separations, interactions with the surrounding environment from DM(through dynamical friction) play a more significant role. 
The frequency of the emitted GWs in the source frame is 
\begin{equation}
    f_s = \frac{\omega}{\pi},
\end{equation}
where the absence of a factor of 2 reflects the spin-2 nature of gravitational waves. For a GW source at redshift \( z \), the observed frequency, $f$ is related through
\begin{equation}
    f = \frac{f_s}{1+z}.
\end{equation}

Two primary mechanisms contribute to the orbital energy loss which are :

     \underline{Gravitational Wave Emission} The power radiated via GW emission is given by  \cite{Maggiore:2007ulw}
    \begin{equation}
        \mathcal{W}_{GW} = \frac{32}{5} G_N \mu^2 \omega^6 r^4,
    \end{equation}
    where \( \mu = \frac{M_1 M_2}{M_1 + M_2} \) represents the reduced mass. This follows from the quadrupole radiation of gravity waves. Systems with larger black holes (considering \( M_1 \sim M_2 \)) dominate this radiation.

     \underline{Dynamical Friction from ULDM}
    Another process of energy loss is the dynamical friction(DF) which occurs when a mass moves in a background of ULDM particles. According to Refs.~\cite{Lancaster:2019mde,Chandrasekhar:1943ys,Yue:2018vtk,Cirelli:2024ssz}, the dynamical friction force exerted on a test mass—here, the reduced mass \( \mu \) of the binary black hole system, by the ULDM background through which it moves, is given by
\begin{equation}\label{DF_formula}
    \vec{F}_{\rm DF} = \frac{4\pi G_N^2 \mu^2 \rho \,\, \vec{v}_{\rm rel}}{v_{\rm rel}^3} \, C(k\,r, \beta)
\end{equation}

    Here, \( v_{rel} \) denotes the relative velocity between the approaching black hole and ULDM particle, and  \( \rho \) is given by Eq.\ref{density_eq}.

\begin{figure}
    \centering
    \includegraphics[width=0.9\linewidth]{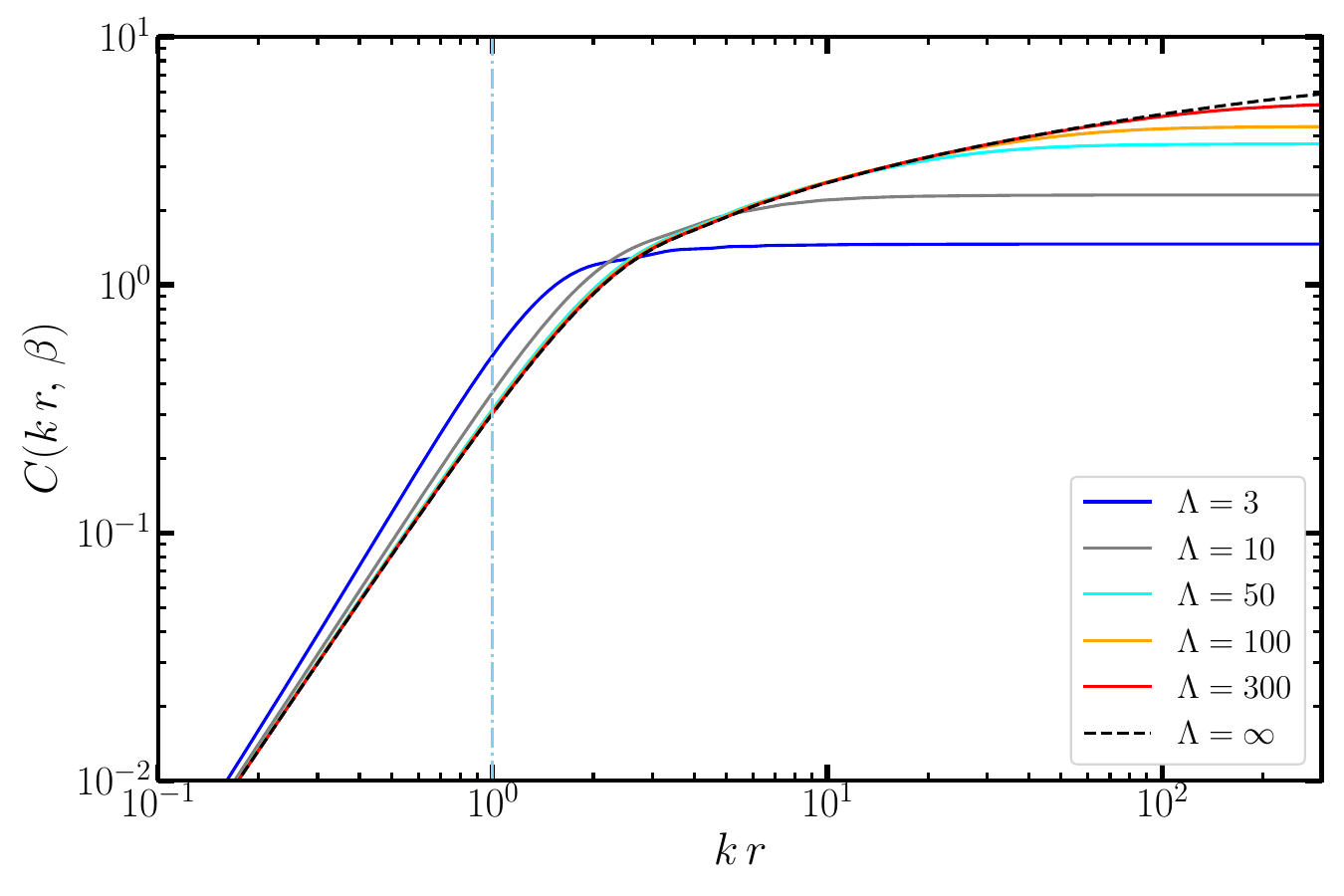}
    \caption{This figure illustrates the dimensionless function \( C(k\,r, \beta) \), which enters the expression for the dynamical friction force experienced by a test object of mass \( \mu \) moving with relative velocity \( v_{\mathrm{rel}} \) through the ULDM~\cite{Hui:2016ltb,Lancaster:2019mde}. The cyan dot-dashed line is drawn at \( k\,r = 1 \), marking the transition between particle-like and wave-like behaviour.}
\label{C_factor_plot}
\end{figure}

   In our case, the relative velocity \(v_{\rm rel}\) between the approaching SMBH and the ULDM particles is comparable to the orbital velocity of the SMBH, i.e., \(v_{\rm rel} \approx \omega r\), for SMBH masses \(\lesssim 10^{9.5}\, \rm M_{\odot}\) and ULDM masses around \(10^{-21}\,\rm eV\), corresponding to the nanohertz frequency range.
 Taking these into consideration,
    the energy dissipation due to this dynamical friction is
    \begin{equation}
        \mathcal{W}_{DF} = \vec{F}_{DF}\,. \,\vec{v}_{rel}= \frac{4\pi G_N^2 \mu^2 \rho}{\omega r} C(k\,r,\beta),
    \end{equation}
The quantity \(C(k\,r,\beta) \) accounts for gravitational feedback effects arising from infrared divergences due to the long-range nature of gravitational interactions. Depending on the values of \( k r \) and \( \beta \), either the wave-like or particle-like character of ULDM becomes relevant. As thoroughly discussed in Appendix D of \cite{Hui:2016ltb}, the general behavior of this correction factor is illustrated in Fig.\ref{C_factor_plot}.
In our analysis, we consider the relevant physical scales that determine the value of \( C \). The parameter \( \beta \) is defined as \( \beta = 2\pi G \mu / (v^2 \lambda) \), where 
the de Broglie wavelength of ULDM, $\lambda$ is given by
\begin{equation}
 \lambda = 2\pi \times 192\, \mathrm{pc} \left( \frac{10^{-21}\, \mathrm{eV}}{m} \right) \left( \frac{10\, \mathrm{km\,s^{-1}}}{v} \right)
\end{equation}
, while the dimensionless quantity \( \Lambda \) is given by \( \Lambda = v^2 r / (G \mu) \). The product \( k r \) can then be estimated as, \( k r = \Lambda \beta \). In order to observe significant wave effects, the condition \( k\,r \lesssim 1 \) must be satisfied. As evident from this analysis, stronger wave effects correspond to smaller values of \( C \), which in turn suppress the efficiency of dynamical friction.

The factor $C(k\,r,\beta)$ depends on the
binary separation $r$ through both $k\,r$ and $\beta$. Using the
definitions of $\beta$ and $\Lambda$, the product $k\,r$ simplifies to
$k\,r = 2\pi r/\lambda_{\rm dB}(v(r))$, with $v(r)$ fixed by Kepler's
law (Eq.~\ref{vel_kepler}). For the benchmark SMBH mass ($M_\bullet=10^{8.5}
M_\odot$), evaluating this over separations from $r=0.01$~pc to
$r=10$~pc gives $k\,r$ ranging from $\approx0.1$ to $\approx2.8$,
bracketing the wave-dominated regime in which the benchmark value
$C\simeq0.5$ has been adopted throughout this work.

To assess the sensitivity of our results to this choice, we note that
the critical radius scales as $r_{\rm cr}\propto C^{-2/11}$ (Eq.~\ref{crit_radius}),
so that the resulting frequency scales as
$f_{\rm cr}\propto C^{3/11}$ (using Kepler's law). Evaluating this scaling across the full
range of $\hat\lambda\in[-1,+1]$ considered in this work, we find that
varying $C$ over $0.1$--$2.7$ (corresponding to $r=0.01$--$10$~pc)
consistently shifts $f_{\rm cr}$ relative to the $C=0.5$ benchmark by a
roughly constant fractional factor: for
$C=0.1$, $f_{\rm cr}$ ranges from $3.6$~nHz to $0.18$~nHz across
$\hat\lambda=-1$ to $+1$, compared to $5.58$~nHz to $0.28$~nHz at
$C=0.5$, and to $8.8$~nHz to $0.45$~nHz at $C=2.8$.

In addition to the dynamical friction induced by a ULDM background, gaseous or dusty environments can also exert significant drag on inspiralling SMBHs\cite{Ostriker:1998fa,2013MNRAS.429.3114C,2023A&A...674A.217L}.
However, in galactic centers and during SMBH inspiral, the motion often becomes transonic or subsonic (i.e. SMBH's velocity relative to the medium is  \(\lesssim 1\)) owing to the high sound speed of the nuclear gas \cite{2005ApJ...623L..67K,Mayer:2007vk,2023A&A...674A.217L}, leading to a reduced dynamical friction efficiency.
Recent cosmological hydrodynamical simulations predict and calibrate DF models in realistic environments, testing both gas and collisionless components self-consistently \cite{2022MNRAS.510..531C}. These studies show that, while gas drag can assist orbital decay, DF from collisionless matter (stars and dark matter) generally dominates beyond the first Gyr of evolution, especially for black holes with masses \(M_{\bullet} \gtrsim 10^{7}\,M_{\odot}\). 
Additionally the environmental effects such as the eccentricity of the orbital motion, gaseous 
circumbinary disk and loss-cone hardening effects onto the SGWB have been briefly discussed in Appendix~\ref{env_degeneracy}.

We now proceed to calculate the GW strain spectrum generated by SMBH merger events occurring across different galaxies and redshifts within ULDM environments. The GW spectrum is characterized by the dimensionless strain amplitude \( h_c(f) \), which is related to the GW energy density spectrum \( \Omega_{\mathrm{GW}}(f) \) via~\cite{Phinney2001APT}:
\begin{equation}\label{GW-Strain-equation}
\begin{split}
\rho_c \, \Omega_{\mathrm{GW}}(f) &= \frac{\pi}{4 G_N} f^{2} h_c^2(f) \\
&= \int \frac{dz}{1+z} \int dX \, \frac{dn_s}{dz\, dX} 
\left( f_s \frac{dE_{\mathrm{GW}}}{df_s} \right) \Bigg|_{f_s = f(1+z)},
\end{split}
\end{equation}

where $dn_{s}/dz dX$ is the number of merger events per unit comoving volume between redshifts $z$ and $z+dz$. X collectively denotes all SMBH sources distributed over those redshifts. Computations related to the SMBH merger rate are given in Appendix~\ref{SMBHB_merger_rate}.

The strain can be expressed as a power law:
\begin{equation}\label{GW_powerlaw}
    h_{c}^{2}(f)= \left[A_{GW}\left(\frac{f}{f_{ref}}\right)^{\beta}\right]^{2}
\end{equation}
Here, $\rho_{c}= 3H_{0}^{2}/8\pi G_{N} $ is the critical density of the universe, and $A_{GW}$ is the amplitude and $f_{ref}$ is the reference frequency at which the PTA measurements provide best accuracy.

Including energy losses from ULDM dynamical friction and GW emission, the GW energy spectrum is\cite{Dror:2021wrl}:
\begin{equation}
    \frac{dE_{GW}}{df_{s}}=\frac{\mu}{3} \Bigg[\pi G_{N}(M_{1}+M_{2})\Bigg]^{2/3} f_{s}^{-1/3} \frac{1}{1+\frac{\mathcal{W}_{DF}}{\mathcal{W}_{GW}}}
\end{equation}
\begin{figure*}[t]
    \centering
    \includegraphics[width=0.49\textwidth]{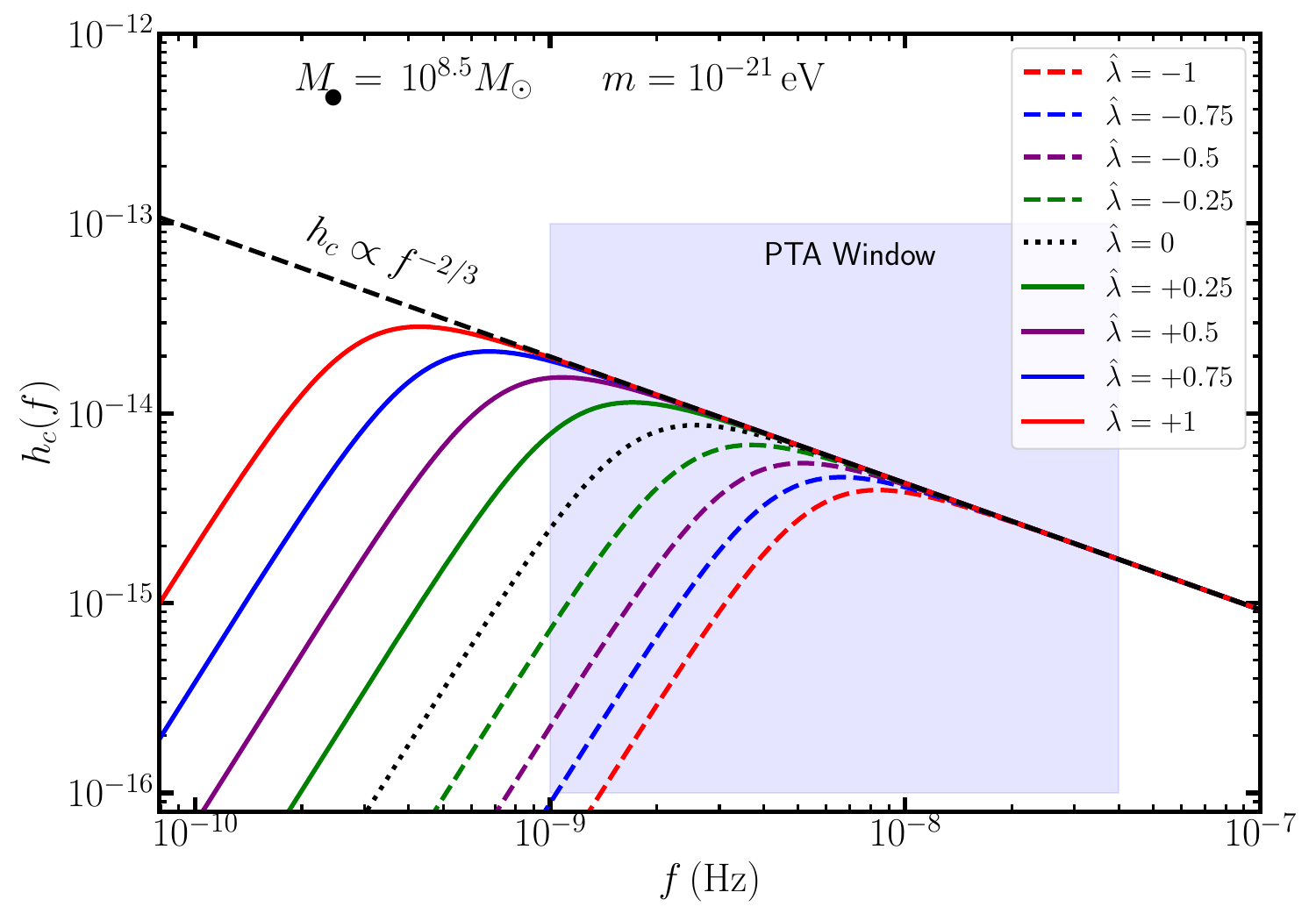} 
    \includegraphics[width=0.49\textwidth]{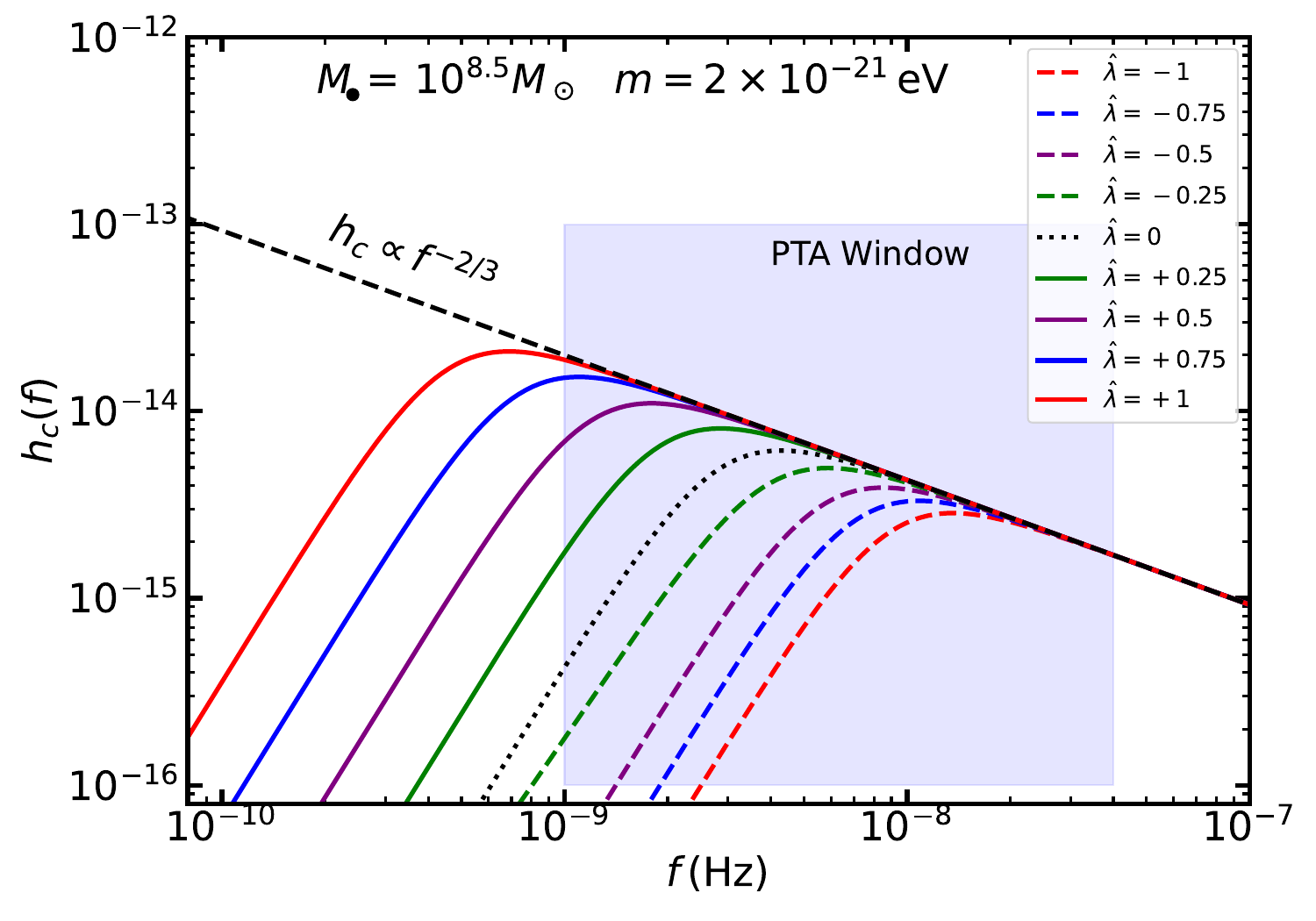}
    \caption{The GW strains corresponding to different DM density profiles, characterized by distinct self-coupling parameters, are displayed. (Left) For a ULDM mass of $10^{-21}\rm eV$ and an SMBH mass of $10^{8.5}\rm M_{\odot}$ of the binary, the strains associated with varying $\hat{\lambda}$ parameters are shown, exhibiting deviations from the power-law behaviour ($h_c\propto f^{-2/3}$).(Right) The strains are presented for a ULDM mass of $2\times10^{-21}\rm eV$ . The blue-shaded region represents the PTA observation window for the SGWB generated by black hole mergers in a solitonic dark matter environment with self-interacting ULDM particles. }
    \label{fig:GW-Strain1}
\end{figure*}
\begin{figure*}[t]
    \centering
    \includegraphics[width=0.49\textwidth]{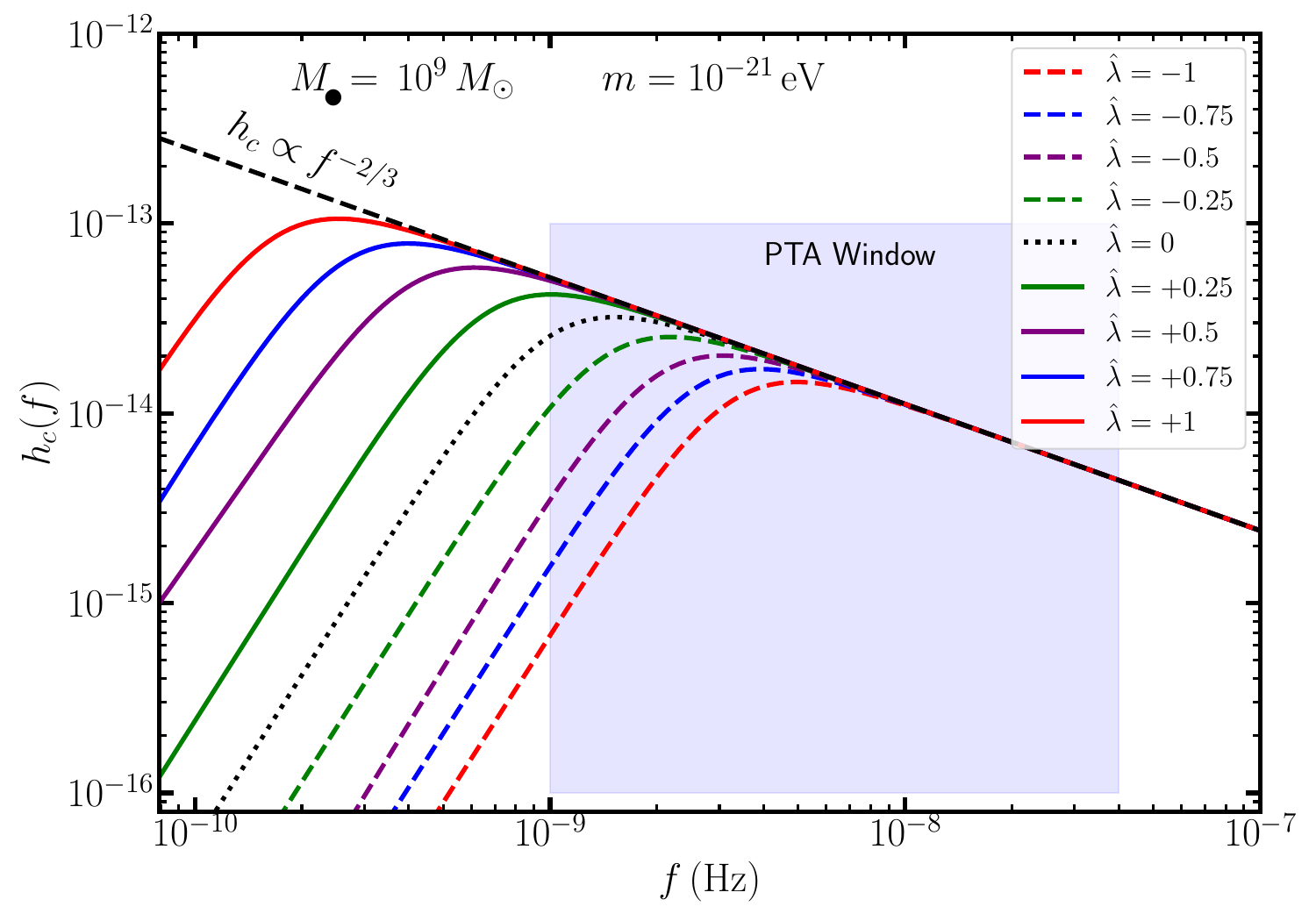}
     \includegraphics[width=0.49\textwidth]{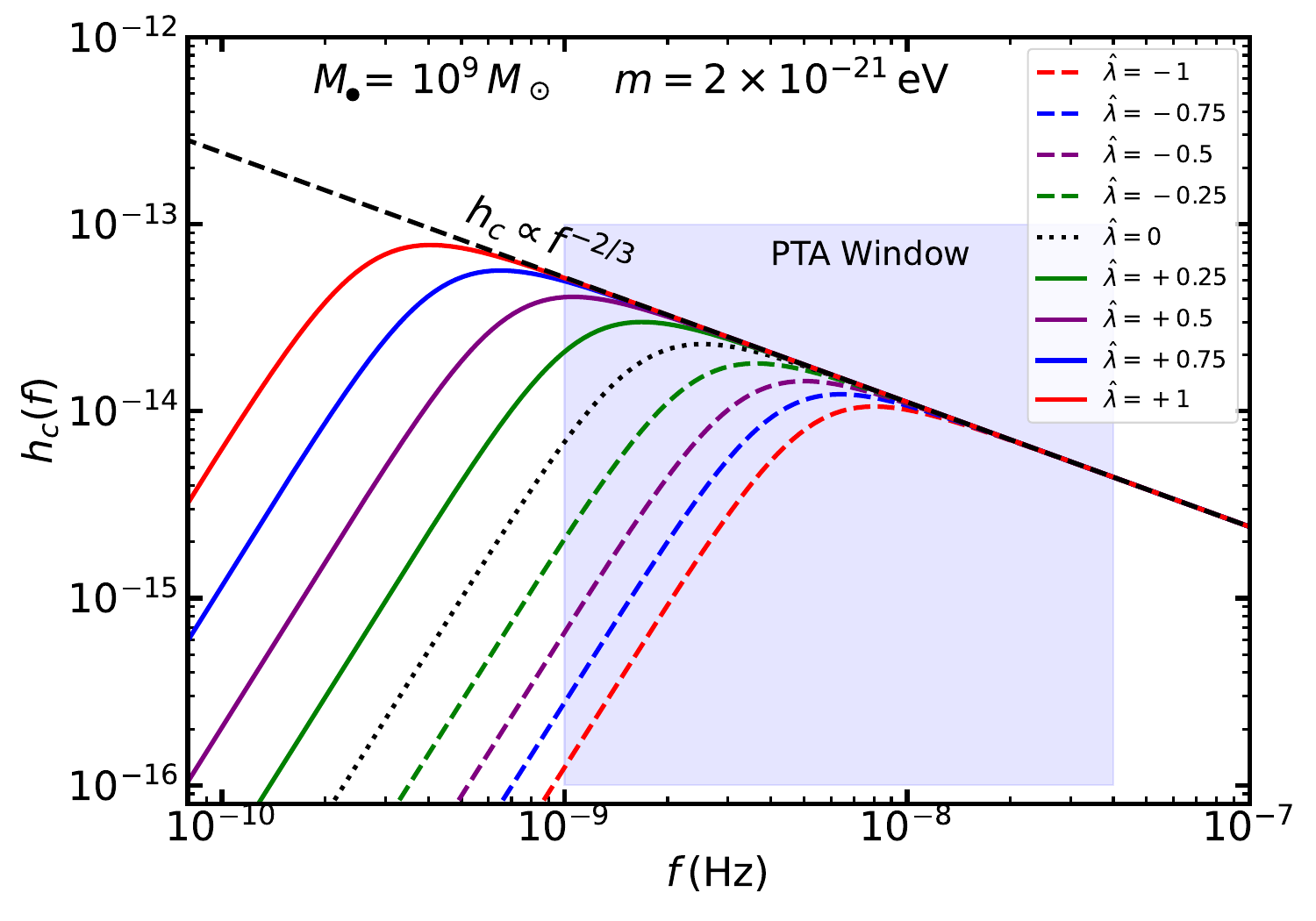}
    \caption{The GW strains corresponding to different DM density profiles, characterized by distinct self-coupling parameters, are displayed. 
    (Left) For a ULDM mass of $10^{-21}\rm eV$ and an SMBH mass of $10^{9}\rm M_{\odot}$ of the binary, the strains associated with varying $\hat{\lambda}$ parameters are shown, demonstrating deviations from the power-law behaviour ($h_c\propto f^{-2/3}$).
    (Right) The strains are shown for a ULDM mass of $2\times10^{-21}\rm eV$. The blue-shaded region marks the PTA observation window for the SGWB .}
    \label{fig:GW-Strain2}
\end{figure*}
The density $\rho$ is determined using the soliton density profiles outlined in Section \ref{Density profiles}. We assume the mass ratio of the merging SMBHs is $q_{\bullet}$, such that $M_{1}+M_{2}=M_{\bullet}(1+q_{\bullet})$, where $M_{1}\equiv M_{\bullet}$ and $q_{\bullet}\lesssim$ 1 in this analysis. Now we identify the distance scale at which the dynamical friction becomes the dominant energy loss mechanism.
We define the critical radius $r_{cr}$ of the soliton core as the distance where the two dissipative forces are equal, i.e., $\mathcal{W}_{GW}=\mathcal{W}_{DF}$. Assuming the central soliton density $\rho_{0}=\rho(r=0)$, the critical radius is expressed as:
 \begin{equation}\label{crit_radius}
    r_{cr}=\sqrt[11]{\frac{64 G_{N}^{5} M_{\bullet}^7 (1+q_{\bullet})^{7}}{25\pi^{2} \rho_{0}^2 C(k\,r,\beta)^2}}
\end{equation}

This expression depends on the soliton density, which itself is a function of the ULDM mass, the self-coupling parameter $\lambda$,  and the mass of the merging SMBH. For a benchmark case where $M_{\bullet}=10^{8.5}M_{\odot}$, $m=10^{-21}$eV and dimensionless self-coupling $\hat{\lambda}_{in}=-1.0 $, the critical radius evaluates to approximately $8\times 10^{-3}$pc. Using Kepler's law, the corresponding radial frequency $f_{cr}$ is 3.72nHz, well within the detection range of current PTA experiments. Putting the loss powers $\mathcal{W}_{GW}$ and $\mathcal{W}_{DF}$, we get the frequency spectra, 
\begin{equation}\label{energy-loss-eqn}
    \frac{dE_{GW}}{df_{s}}=\frac{\pi^{2/3}q_{\bullet}G_{N}^{2/3}M_{\bullet}^{5/3}f_{s}^{-1/3}}{3(1+q_{\bullet})^{1/3}}\frac{1}{1+\left(\frac{f_{cr}}{f_{s}}\right)^{11/3} e^{-\left(\frac{f_{sol}}{f_{s}}\right)^{2/3}}}
\end{equation}
From the above expression, it is evident that in the absence of $\mathcal{W}_{DF}$, the energy loss follows the standard power-law behaviour, $dE_{GW}/df_{s}\propto f_{s}^{-1/3}$. However, the presence of dynamical friction, influenced by the soliton dark matter density, distorts the strain amplitude. The radial frequency $f_{sol}$ corresponds to the soliton radius $r_{sol}$ of the dark matter density. When $f_{cr}< f_{sol}$, i.e. the critical radius is greater than the soliton radius the energy loss due to GW emission is more prominent than the dynamical friction. In this case the spectrum follows the standard power-law behaviour. On the other hand, when the critical radius is inside the soliton core ($f_{cr}> f_{sol}$), the dynamical friction is dominant and the frequency spectrum follows a completely different frequency dependence. Due to the soliton core made of the ULDM, the strain of the spectrum shows a sudden dip. 

The depth of this
dip depends on the relative location of the two characteristic frequencies
entering Eq.~\ref{energy-loss-eqn}: $f_{\rm cr}$, at which the
dynamical-friction and GW-emission powers balance ($W_{\rm DF}=W_{\rm GW}$),
and $f_{\rm sol}$, which marks the frequency corresponding to the outer edge
of the soliton core beyond which the ULDM density falls off exponentially.
When $f_{\rm cr}$ and $f_{\rm sol}$ lie close to one another, the onset of
dynamical-friction domination coincides with the region where the available
soliton density is itself declining steeply, so the two suppression effects
compound rather than acting over separate frequency ranges, producing a
deeper and narrower dip in the strain spectrum.

\section{Results}\label{results}
\subsection{The GW spectrum}
Using Eq.~\ref{GW-Strain-equation} and substituting the energy loss power from Eq.~\ref{energy-loss-eqn}, we derive the strain of GW. The soliton density profile influences this spectrum through dynamical friction, as discussed in the previous section.
The strains are shown in Fig.~\ref{fig:GW-Strain1} for the merging SMBH masses $10^{8.5}\rm M_{\odot}$ and in Fig.~\ref{fig:GW-Strain2} for merging SMBH masses $10^{9}\rm M_{\odot}$. The cases for two ULDM masses are shown in sub-figures.

In the absence of self-interactions among ULDM particles, represented by $\hat{\lambda} = 0$, the GW spectrum follows the black dotted line. Non-zero values of $\hat{\lambda}$, which denote self-interactions, modify the spectrum. Specifically, positive $\hat{\lambda}$, corresponding to repulsive soliton cores, are depicted with solid lines, while negative $\hat{\lambda}$, representing attractive soliton cores, are shown with dashed lines. If GWs are generated purely by mergers without incorporating the effects of dynamical friction (i.e., ignoring dark matter densities), the GW strain follows a simple power-law behaviour ($h_{c} \propto f^{-2/3}$), represented by the black dashed line. The slope of the spectrum deviates from this power-law due to the influence of dark matter density, where the soliton serves as the dark matter profile in this analysis. 

The PTA experiments operate in the frequency range of $1$--$30$~nHz and can probe features of the ULDM, such as self-interactions. Within this window, PTA observations can place qualitative limits on the value of $\hat{\lambda}$. 
This qualitative limits would carry characteristic signatures detectable through the GW spectrum.

\subsection{Typical values of self-coupling $\hat{\lambda}$}\label{Theoretical_lambda}
The interplay between the strength of gravity and the self-interaction of ULDM particles plays a crucial role in the formation of soliton cores. From the GPP equations, the ratio of the gravitational potential energy, $m \Phi \Psi$, to the self-interaction energy, $\frac{\lambda}{8m^{3}} |\Psi|^{3}$, is given by:
\begin{equation}
    \frac{\text{gravity}}{\text{self-interaction}} = \frac{8 m^{4} \Phi}{\lambda |\Psi|^{2}} 
    = \frac{4}{\lambda}\left(\frac{M_\text{pl}}{m}\right)^{-2} \left(\frac{L}{m^{-1}}\right)^{2},
\end{equation}
L being the physical size of the soliton.
This shows that gravity becomes more significant than self-interactions as the soliton size increases. Using the definition of the dimensionless self-coupling parameter $\hat{\lambda}$, this ratio can be rewritten as:
\begin{equation}
    \frac{\text{gravity}}{\text{self-interaction}} = \frac{1}{2\hat{\lambda}}.
\end{equation}
For gravity to dominate over self-interactions, we require:
\begin{equation}\label{th_lambda_limit}
    -\frac{1}{2} \leq \hat{\lambda} \leq \frac{1}{2}.
\end{equation}

Here, a negative $\hat{\lambda}$ corresponds to attractive self-interactions, while a positive $\hat{\lambda}$ corresponds to repulsive self-interactions. In this study, we do not focus on cases where self-interactions dominate gravity, as the GPP equations enter into a highly non-linear regime, introducing additional complexities.

Eq.~\ref{th_lambda_limit} gives a dimensionless limit of the self-coupling parameter $\hat{\lambda}$, as long as gravity dominates over the self-interaction of ULDM particles and that the soliton cores at the galactic centers remain stable.
\begin{figure*}[t]
    \centering
    \includegraphics[width=0.49\textwidth]{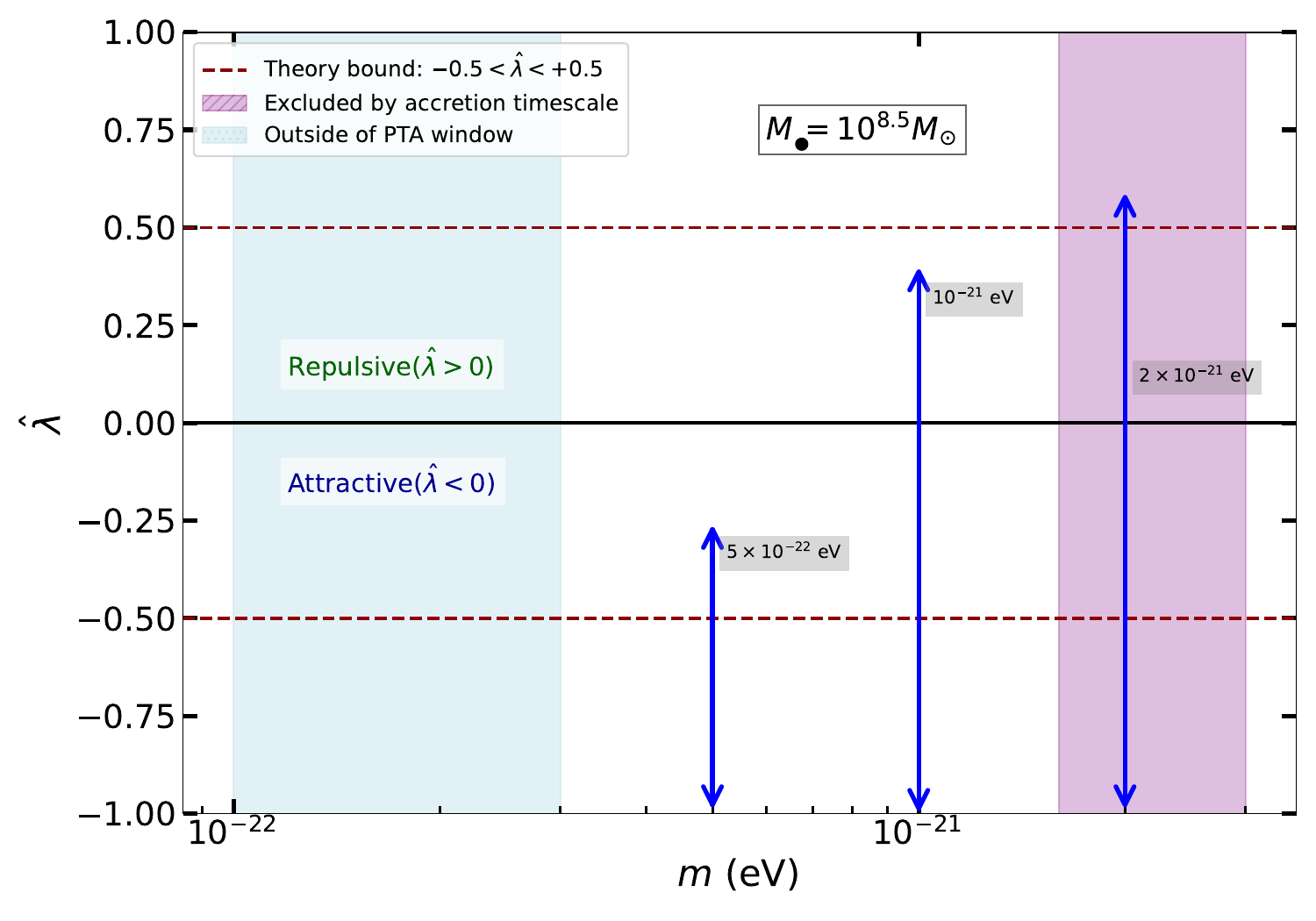}
    \includegraphics[width=0.49\textwidth]{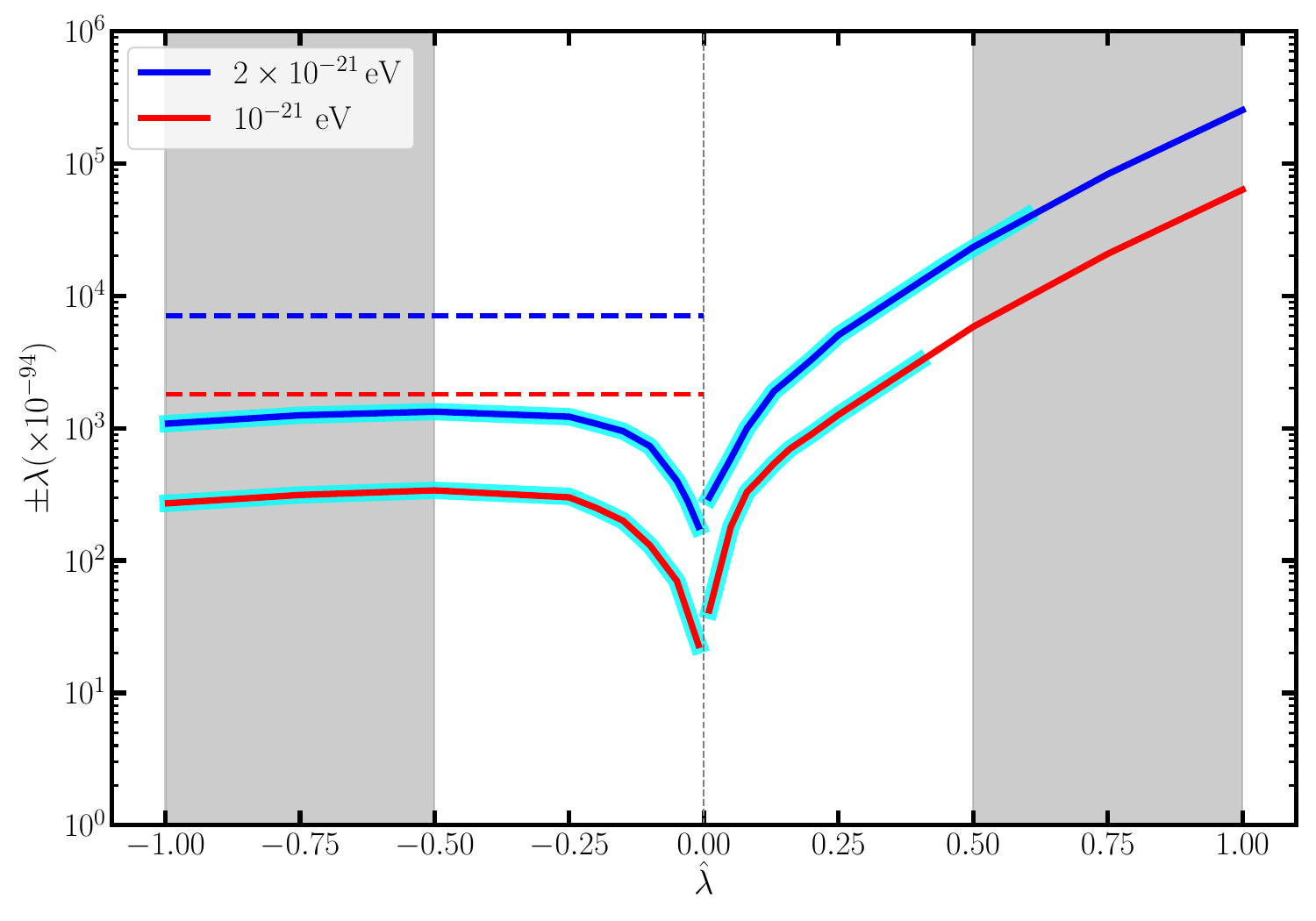}
    
    \caption{{\it Left:}  Qualitative limits on the parameter space $(m, \hat{\lambda})$. The sign of $\hat{\lambda}$ indicates whether attractive($\hat{\lambda}<0$) or a repulsive coupling($\hat{\lambda}>0$). Horizontal red dashed lines show the bound $-0.5 < \hat{\lambda} < +0.5$  based on theoretical considerations discussed in Section.~\ref{Theoretical_lambda} . The region $m > 1.6\times10^{-21}$ eV is excluded by accretion timescale limits( using Eq.~\ref{accr_time}), while $m < 3\times10^{-22}$ eV lies outside the PTA sensitivity window. The blue arrows signify the region over which PTA can probe the $\hat{\lambda}$ parameter for 3 different ULDM masses. 
    {\it Right:} The estimated values of the  $\lambda$ parameter are shown along with the dimensionless $\hat{\lambda}$, considering both positive and negative values(-ve estimate of $\lambda$ is for the -ve values of the $\hat{\lambda}$ and positive $\lambda$ is for the positive $\hat{\lambda}$ values). The plots are presented for two different ULDM masses but keeping the SMBH mass to $10^{8.5}\rm M_{\odot}$. The cyan shading over the plots represents the range that can be probed through PTA observations, while the gray shaded region indicates the excluded region based on theoretical considerations. Dashed lines(blue and red), corresponding to the attractive self-coupling parameters derived using Eq.~\ref{limit_lambda_attractive_pot}, illustrate the upper limit on the strength of $\lambda$.}
    \label{fig:self-coupling}
\end{figure*}

\subsection{Allowed values of $\lambda$}
  The typical frequency band accessible to PTA experiments for the SGWB lies in the $1$--$30\,\mathrm{nHz}$ range. Within this window, a sharp dip in the strain spectrum can constrain the self-coupling parameter $\hat{\lambda}$, which sets the dimensionful coupling strength $\lambda$. For an equal-mass binary with a black hole mass of $10^{8.5} M_{\odot}$, attractive self-coupling values ($0$ to $-1$) fall inside the PTA band when the ULDM mass is $10^{-21}\,\mathrm{eV}$. For lighter ULDM masses, the spectral peak shifts to lower frequencies, leaving only negative $\hat{\lambda}$ values within the probed region, as illustrated in the left panel of Fig.~\ref{fig:self-coupling}.  

  From theoretical considerations---namely that black hole gravity dominates over soliton self-gravity---the maximum allowed value of $\hat{\lambda}$ is restricted to $\pm 0.5$. This theoretical bound, shown as the red dashed lines in the left panel of Fig.~\ref{fig:self-coupling}, limits the extent to which the SGWB can constrain the coupling. In addition, the purple shaded region marks the parameter space excluded by the accretion timescale constraint according to Eq.~\ref{accr_time}.

From Eq.~\ref{eq:self_coupling}, the strength of the coupling for a ULDM mass of \(10^{-21}\) eV is approximately \(\mathcal{O}(10^{-96})\) for both attractive and repulsive self-couplings. The \(\hat{\lambda}\) parameter scales with the $s^{2}$ factor, increasing the coupling strength by several orders of magnitude. For instance, for a ULDM mass of \(10^{-21}\) eV and \(\hat{\lambda} = -0.5\), the coupling strength \(\lambda\) increases to approximately \(-3.4 \times 10^{-92}\). Similarly, for a ULDM mass of \(2\times10^{-21}\) eV and \(\hat{\lambda} = -0.5\), \(\lambda\) is approximately \(-1.33 \times 10^{-91}\).

As the scaling $s$ increases, the strength of \(\lambda\) grows even larger for positive \(\hat{\lambda}\) values. 
However, for a ULDM particle mass of ($10^{-21},\mathrm{eV}$), repulsive self-coupling (\(+\hat{\lambda}\)) in the range ($0 \leq \hat{\lambda} \leq 0.4$) falls within the observational window. For a heavier ULDM particle mass of ($2\times10^{-21},\mathrm{eV}$), the entire self-coupling range considered in this work, ($-1 \leq \hat{\lambda} \leq 0.6$), lies within the observational window. In contrast, for lighter ULDM particle masses, the parameter space corresponding to repulsive self-interactions shifts outside the observational window, as shown in Fig.~\ref{fig:GW-Strain1} and Fig.~\ref{fig:GW-Strain2}. Furthermore, increasing the masses of the SMBHs in the binary population changes the amplitude of the SGWB while leaving its spectral slope essentially unchanged.

For large strengths of attractive self-interactions, the balance between gravity, self-interactions, and quantum pressure cannot be maintained, resulting in the absence of stable soliton solutions. A stable soliton exists only up to a maximum possible soliton mass, \(M_c\), given by\cite{Chavanis:2011zi,Levkov:2016rkk}:

\begin{equation}
    M_c \approx 10.2 \frac{M_\text{pl}}{(-\lambda)^{1/2}}
\end{equation}

where the negative sign indicates an attractive self-interaction. For a galaxy with a mass \(M_\text{halo} = 10^{13} M_\odot\), the soliton mass can be determined using Eq.~\ref{dimension_less_mass} and then incorporating an appropriate scaling. From this, an upper limit is established to ensure the soliton mass does not exceed \(M_c\), given by:

\begin{equation}\label{limit_lambda_attractive_pot}
    |-\lambda| \leq 1.77 \times 10^{-91} \left(\frac{m}{10^{-21}}\right)^2
\end{equation}

In the right panel of Fig.~\ref{fig:self-coupling}, the maximum possible coupling is represented by dashed lines corresponding to different masses of ULDM and an SMBH mass of $10^{8.5}\rm M_{\odot}$ in binary populations having equal masses. The cyan shading over the blue and red plots delineates the range of self-coupling limits which can be probed from the SGWB spectrum amplitude.

\subsection{Relation between the GW amplitude and the $\gamma_{GW}$ parameter}
\begin{figure*}
    \centering
     \includegraphics[width=0.49\textwidth]{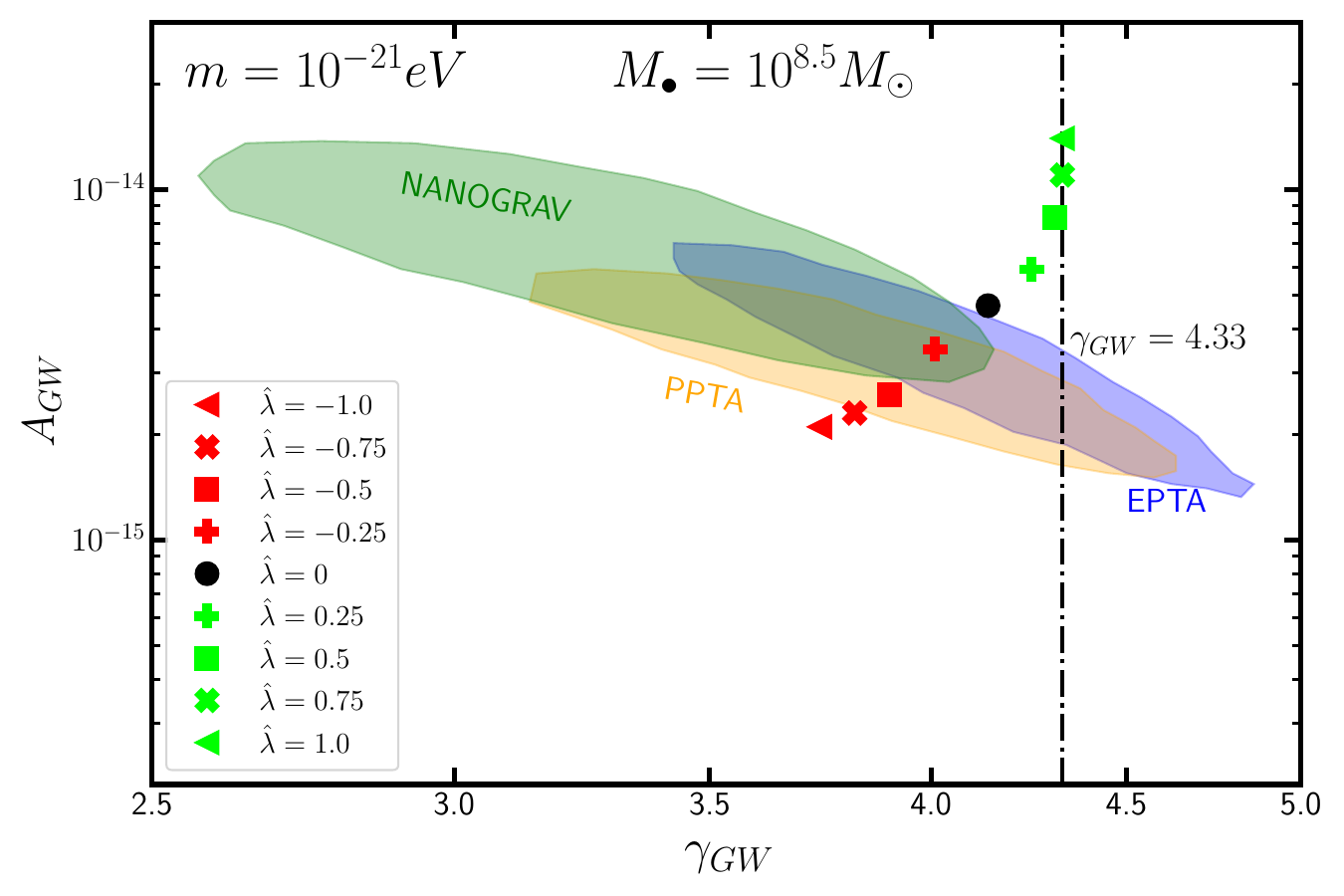}
    \includegraphics[width=0.49\textwidth]{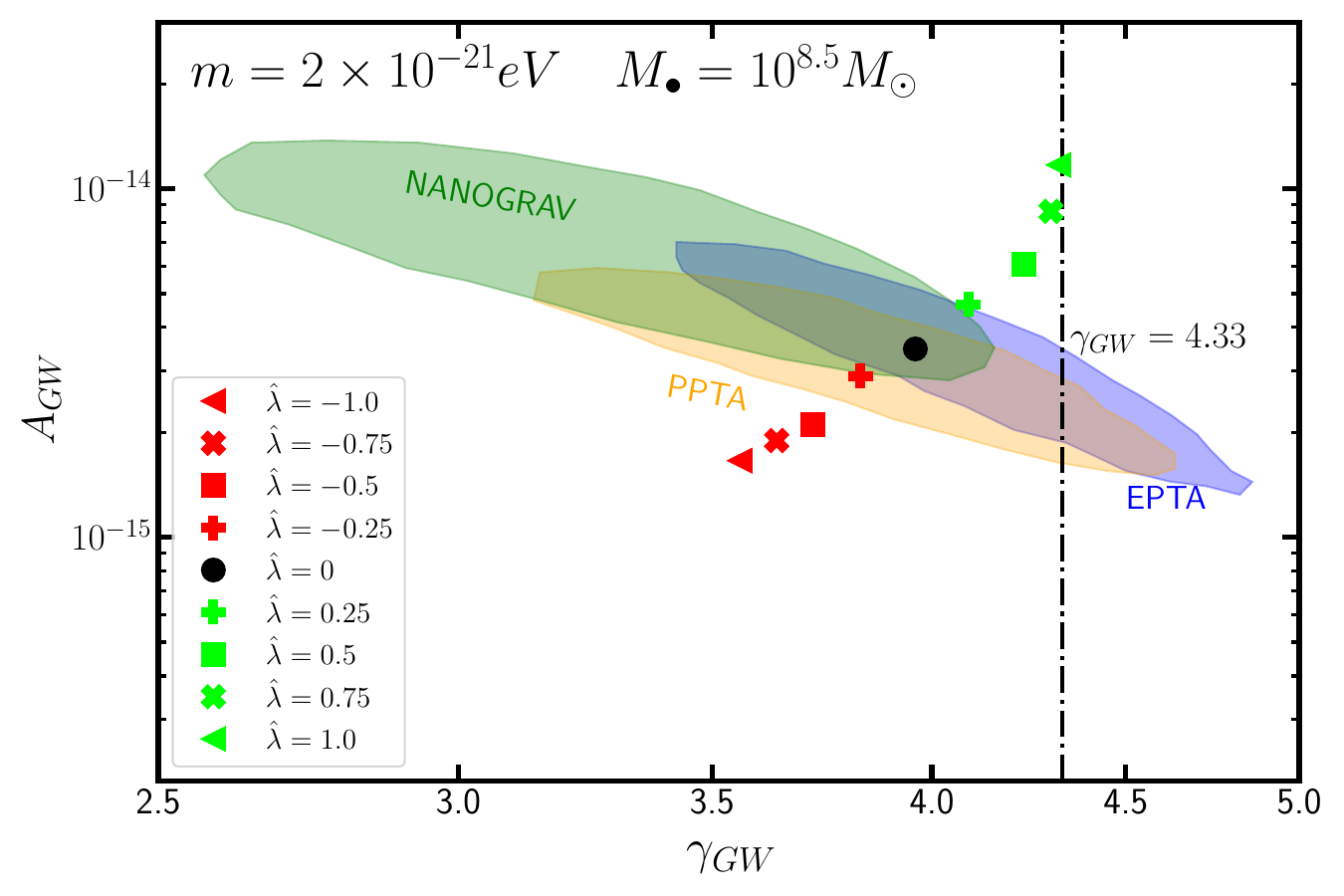}
    \caption{ The amplitude of the GW strain with change of $\gamma_{\rm GW}$ parameter for PTA observations by NANOGRAV, EPTA, and PPTA are represented by the shaded regions\cite{DeRocco:2023qae} within 2$\sigma$ confidence interval for the GW signal. For a reference frequency ($f_{\rm PTA}$) of 1 nHz, the amplitude and $\gamma_{\rm GW}$ values corresponding to different $\hat{\lambda}$ parameters are depicted using various legends. For a equal mass non-eccentric binary systems, keeping the SMBH mass fixed at $10^{8.5}\,\mathrm{M_{\odot}}$, the correlation between $A_{\rm GW}$ and $\gamma_{\rm GW}$ for a ULDM mass of $10^{-21}\,\mathrm{eV}$ is shown in the left figure, while for a ULDM mass of $2\times10^{-21}\,\mathrm{eV}$, it is displayed in the right figure. The dashed vertical line at the $\gamma_{GW}=4.33$ is the corresponding value for a GW strain that exactly follows the power-law $h_c\propto f^{-2/3}$.    }
    \label{fig:A_vs_gamma}
\end{figure*}
PTAs probe the SGWB generated by SMBH mergers in the nHz frequency range. The PTA data analyzes the strain spectrum, typically parameterized as a power law: $h_{c}(f)=A_{GW}(f/f_{PTA})^{\beta}$
where \( f_{\text{PTA}} \) is a reference frequency. The GW spectrum is characterized by the amplitude \( A_{GW} \) and the slope parameter \( \gamma_{GW} = 3 - 2\beta \).

Astrophysical environments in the vicinity of SMBHs can alter the spectral slope, causing it to deviate from that expected solely from gravitational wave radiation losses.
 In circular binaries driven primarily by GW radiation losses, the spectral index is \( \beta = -2/3 \), corresponding to \( \gamma_{GW} = 13/3 \). However, if dynamical friction dominates, deviations in \( \gamma_{GW} \) arise, providing a pathway to limit the self-coupling parameter of ULDM.

	Tentative SGWB signal found at frequencies above 1 nHz as fit by NANOGrav,
EPTA, and PPTA is often expressed through correlation of the GW amplitude $A_{GW}$ and the $\gamma_{GW}$ parameter\cite{DeRocco:2023qae}. So we set the reference frequency at $f_{\text{PTA}} =1\, \text{nHz}$ for this analysis.

The Fig.~\ref{fig:A_vs_gamma} describes the relation between the GW amplitude and the slope parameter $\gamma_{GW}$ for the two cases : the left one is for ULDM mass $m=10^{-21}$eV and the right one is for  $m=2\times10^{-21}$eV and keeping the SMBH mass fixed at $10^{8.5}\rm M_{\odot}$. The effect of the self-coupling, specified by different legends, are shown in the figures. As depicted, the range of the $\hat{\lambda}$ parameter that lies within the correlated regions defined by the current PTA observations is relatively narrow. The cases for SMBH mass $10^{9}\rm M_{\odot}$ have been stacked in the \ref{GW_A_vs_gamma} with the present case for comparison.

\section{Discussion and Conclusion}\label{Discussion}

In this study, we have explored the possibility of placing qualitative limits
on the ULDM self-coupling $\lambda$ using recent PTA observations.
ULDM with masses $\sim(10^{-22}-10^{-20})$~eV can form solitonic cores 
at the centers of galaxies, and the inspiral of black holes within these 
dense solitonic environments can imprint signatures on the stochastic 
gravitational wave background.
The presence of self-interactions among the ULDM particles modifies the 
GW strain through dynamical friction, whose efficiency 
depends sensitively on the solitonic density profile.

We have identified which part of the SMBH population contributes most 
significantly to the GW signal in the nanohertz band.
For non-eccentric, equal-mass SMBH binaries, masses around 
$M_{\bullet}\sim 10^{8.5}\,M_{\odot}$ dominate the strain $h_c$ at 
nanohertz frequencies.
A key finding of this work is that producing observable 
signatures in the PTA band requires a careful fine-tuning among three 
interconnected quantities: the host galaxy halo mass, the inspiralling 
SMBH mass, and the soliton mass formed by ULDM particles of a given 
mass $m$ and self-coupling $\lambda$.
Specifically, the halo mass sets the SMBH mass through the black 
hole--halo mass relation, which in turn determines the soliton mass 
via the soliton--halo mass relation.
Only when these three quantities are mutually consistent with 
$M_{\rm halo}\sim 10^{13}\,M_{\odot}$, $M_{\bullet}\sim 
10^{8.5}\,M_{\odot}$, and $m\sim(3\times10^{-22} - 
2\times10^{-21})$~eV,  does the soliton-induced dynamical friction 
produce a suppression feature that falls within the nanohertz 
sensitivity window of current PTAs.
Departures from this configuration, either through a larger halo mass 
that pushes the SMBH mass beyond $\sim 10^9\,M_{\odot}$, or through 
a ULDM mass outside this range, shift the spectral feature outside 
the observable band or render the soliton unstable against accretion 
by the central SMBH.

The empirical relations between the soliton mass, the central SMBH 
mass, and the halo mass motivate our choice of 
$M_{\rm halo} = 10^{13}\,M_{\odot}$ as the fiducial halo mass for 
solving the GPP equations numerically.
For each solution, the self-coupling parameter $\lambda$ is fixed, 
as shown in Fig.~\ref{fig:GPP-dimensionless}, and the corresponding 
dark matter density profiles are obtained by applying the appropriate 
scaling relations to convert the dimensionless solutions into physical 
units.
The presence of self-coupling $\lambda$ with its characteristic 
signature modifies the DM profiles as shown in 
Fig.~\ref{fig:Density_profile}, with these individual ULDM density 
profiles leading to distinct GW strains presented in 
Fig.~\ref{fig:GW-Strain1} and Fig.~\ref{fig:GW-Strain2}.

The nanohertz window of the spectrum is particularly sensitive to 
recent observations from EPTA, PPTA, and NANOGrav.
Deviations in the strain caused by $\lambda$ from the standard 
power-law behaviour ($h_c\propto f^{-2/3}$) identify the range and 
characteristic signature of $\lambda$ accessible within this frequency 
band.
Based on this analysis, we place qualitative limits on $\lambda$ for specific 
ULDM masses, as shown in Fig.~\ref{fig:self-coupling} for a fixed 
SMBH mass.
Our results indicate that attractive self-coupling ($\lambda < 0$) 
can be best probed through GW strains when ULDM masses lie in the 
range $\sim(3\times10^{-22} - 2\times10^{-21})$~eV, given an SMBH 
mass of $10^{8.5}\,M_{\odot}$.
In contrast, repulsive self-coupling ($\lambda > 0$) can be probed 
within a specific range of the dimensionless $\hat{\lambda}$ parameter 
for ULDM masses around $\sim(10^{-21} - 2\times10^{-21})$~eV.
Soliton cores composed of ULDM particles with masses exceeding 
$\gtrsim 1.6\times10^{-21}$~eV are excluded by accretion timescale 
constraints arising from the gravitational attraction of the central 
SMBH.
Taken together, these results highlight that probing 
ULDM self-coupling through the SGWB requires a precise and 
self-consistent matching of the ULDM particle mass, the self-coupling 
strength, the soliton mass, and the masses of the merging SMBHs,  
a degeneracy that future PTA analyses with larger datasets and improved 
sensitivity will be better placed to break.

Upcoming PTA experiments, such as IPTA30~\cite{Kaiser:2020tlg}, are 
expected to reach frequencies up to $\mathcal{O}(100)$~nHz and 
analyze SMBH populations with masses as low as 
$M_{\bullet}\simeq 10^{6}\,M_{\odot}$~\cite{Ellis:2023owy}, providing 
a more rigorous probe of ULDM self-interactions.
In addition to PTA detection of GWs from SMBHs orbiting soliton cores, 
the planned LISA mission, operating in the milli-hertz frequency band, 
may also provide complementary detection 
capabilities~\cite{LISA:2024hlh}.

\textbf{Acknowledgments:}
We sincerely thank Soumitra SenGupta, Satyanarayan Mukhopadhyay, and Sumanta Chakraborty for engaging discussions. 
In particular, we are indebted to Sumanta Chakraborty for his detailed and constructive feedback on the manuscript. 
We also thank Sourov Roy for carefully going through the draft and offering helpful suggestions. The author acknowledges financial support from the University Grants Commission, Government of India, as a Senior Research Fellow.

\section*{Data and Code Availability}
The data cannot be made publicly available upon publication because they are not available in a format that is sufficiently accessible or reusable by other researchers. The data that support the findings of
this study are available upon reasonable request from the authors.

\appendix


\section{Amount of scaling $s$}\label{Appd_scaling}
Fixing the halo mass, \( M_{\text{halo}} \), there exists an empirical relationship between soliton mass \( M_{\text{sol}} \) and halo mass\cite{Schive:2014hza}:
\begin{equation}
    M_{\text{sol}} = 1.25 \times 10^{9} \left( \frac{M_{\text{halo}}}{10^{12} M_{\odot}} \right)^{1/3} 
    \left( \frac{m}{10^{-22} \, \text{eV}} \right)^{-1} M_{\odot}.
\end{equation}

In this work, we assume \( M_{\text{halo}} = 10^{13} M_{\odot} \) and use the \( \hat{\alpha} \) values from \cite{Davies:2019wgi} to construct density profiles. These values of $\hat{\alpha}$ were obtained using numerical simulation considering virialized halo with SMBH placed at the centers of the galaxies. Table~\ref{tab:example_table} summarizes \( \hat{\alpha} \) and the corresponding dimensionless soliton mass \( \hat{M} \) for different ULDM masses.

\begin{table}[h!]
    \centering
    \begin{tabular}{cccc}
    \hline
    \( M_{\text{halo}} \) & \( m \, (\text{eV}) \) & \( \hat{\alpha} \) & \( \hat{M} \) \\
    \hline
    \( 10^{13} M_{\odot} \) & \( 10^{-21} \) & 0.48 & 0.84 \\
    \( 10^{13} M_{\odot} \) & \( 2\times 10^{-21} \) & 0.55 & 0.75 \\
    \hline
    \end{tabular}
    \caption{Values of \( \hat{\alpha} \) chosen to produce density profiles of halo mass \( 10^{13} M_{\odot} \) without the \( \lambda \) parameter, along with the dimensionless soliton mass \( \hat{M} \) for different ULDM masses.}
    \label{tab:example_table}
\end{table}

With fixed \( \hat{\alpha} \) and nonzero \( \hat{\lambda} \), we compute \( \hat{M} \). For attractive self-interaction (\( \hat{\lambda} < 0 \)) and repulsive (\( \hat{\lambda} > 0 \)), \( \hat{M} \) relates to \( M_{\text{sol}} \) by:
\begin{equation}
    \hat{M} \equiv \frac{G_N M_{\text{sol}} m}{\hbar c}.
\end{equation}

Scaling modifies \( \hat{M} \) as:
\begin{equation}
    \hat{M} \to \frac{1}{s}\hat{M}
\end{equation}

\begin{figure}[h!]
    \centering
    \includegraphics[width=0.9\linewidth]{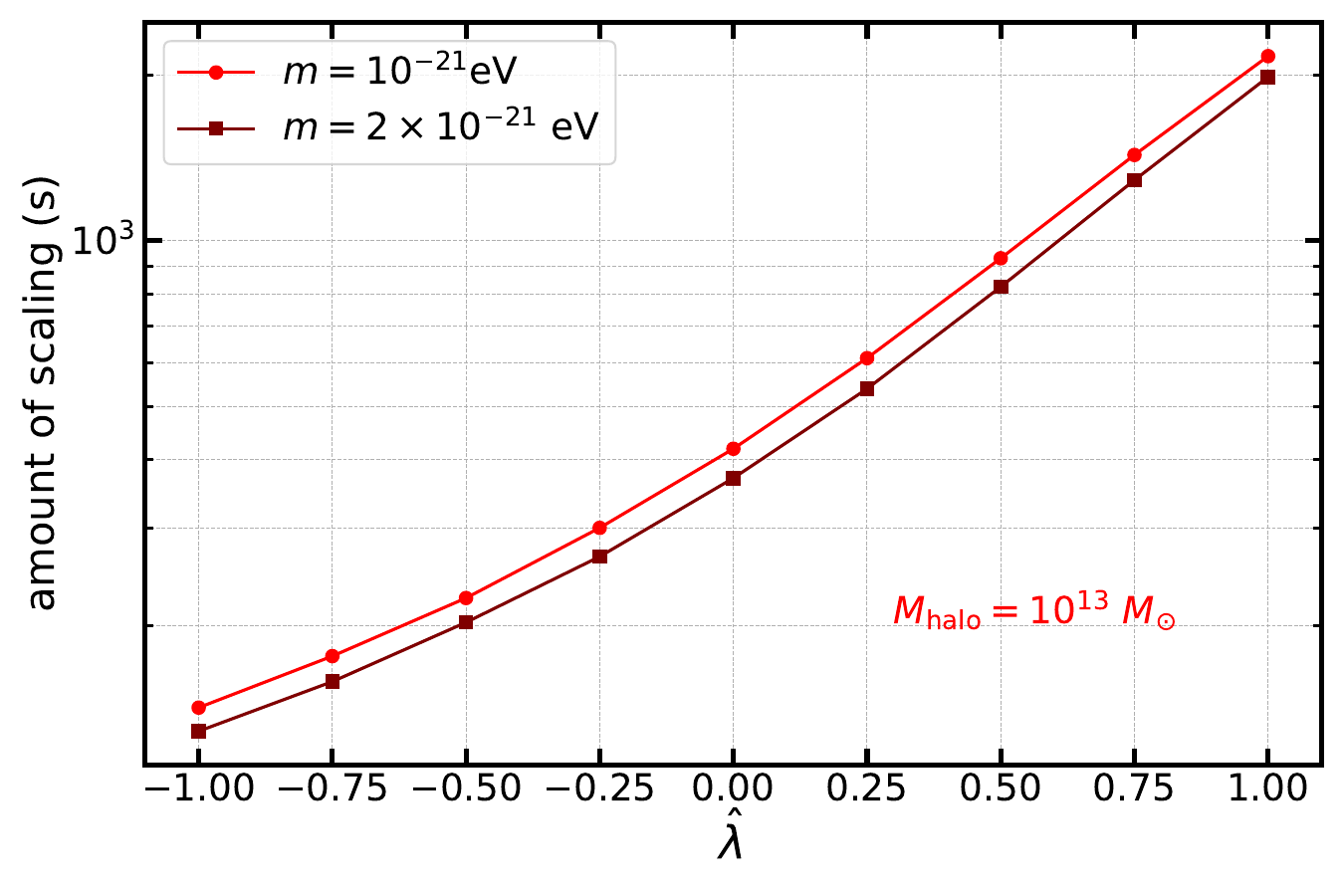}
    \caption{Amount of scaling with different values of \( \hat{\lambda} \) parameter for a fixed halo mass. Three cases are shown for three ULDM masses.}
    \label{fig:scaling-vs-lambda}
\end{figure}

Using the relations above, we calculate the scaling factor \( s \):
\begin{equation}
    s = \frac{1.33 \times 10^{12}}{1.25 \times 10^{9}} 
    \left( \frac{M_{\text{halo}}}{10^{12} M_{\odot}} \right)^{-1/3} \hat{M}.
\end{equation}

The dimensionless soliton mass $\hat{M}$ is calculated numerically through the integral \ref{dimension_less_mass} for each soliton solution with the $\hat{\lambda}$ parameters. 

Figure~\ref{fig:scaling-vs-lambda} illustrates how \( s \) changes with \( \hat{\lambda} \). The scaling \( s \) increases with decreasing ULDM mass for a fixed \( \hat{\lambda} \), and \( s \) also grows as \( \hat{\lambda} \) shifts from negative to positive values.

\section{Environmental degeneracy}\label{env_degeneracy}
Our analysis is restricted to circular, equal-mass SMBH binaries embedded in an otherwise pristine ULDM environment. In realistic systems,
several standard astrophysical mechanisms can also reshape the SGWB
spectrum away from the fiducial $h_c\propto f^{-2/3}$ power law, and we
discuss below a few such environmental effects.

Orbital eccentricity redistributes GW power into higher harmonics and
can flatten or invert the spectrum at low frequencies~\cite{Taylor:2016ftv,Kelley:2017lek}. However, detailed population studies
find that the stellar-scattering hardening rate itself depends only
weakly on eccentricity ($\sim10\%$ variation between $e=0$ and
$e=0.9$), and that binaries evolving through gas-poor mergers tend to
circularize substantially by the time they enter the PTA band, while
gas-coupled binaries settle toward a modest equilibrium eccentricity of
order $e\sim0.5$ \cite{Sesana:2006xw,Kelley:2017lek,2023MNRAS.522.2707S}. Such
eccentricity-driven effects reshape the spectrum smoothly and
gradually, in contrast to the sharp, localized dip predicted by
soliton-induced dynamical friction, whose location is fixed by the
parameters such as $m$ and $\lambda$.

Loss-cone hardening is the phase of SMBH binary inspiral during which stars scattered into the binary's loss cone extract orbital energy and angular momentum through three-body interactions, causing the binary to shrink. Stellar loss-cone hardening and gaseous circumbinary disk
interactions are the two leading candidate mechanisms invoked to
explain the observed low-frequency turnover in the SGWB~\cite{2013CQGra..30x4005M}. Notably, a
recent multi-messenger analysis combining core-scouring observations in
elliptical galaxies with the NANOGrav 15-year dataset finds that
stellar scattering alone -- even under a full loss-cone and a
hardening rate calibrated to be $\sim60\%$ faster than standard N-body
predictions -- is insufficient to reproduce the observed turnover,
strongly implicating gas dynamics as a necessary additional ingredient~\cite{2024MNRAS.528....1H}. 
It demonstrates that the dominant standard
astrophysical explanation for a low-frequency SGWB turnover is, on its
own, already in tension with observations, so any additional
suppression feature attributed to the standard hardening channels
alone cannot trivially reproduce or mimic the localized signature we
predict from ULDM dynamical friction. It underscores that
robustly disentangling our proposed feature from gas-driven or
eccentricity-driven modifications will ultimately require a joint,
multi-mechanism fit to PTA data that includes ULDM dynamical friction
alongside circumbinary disk torques, loss-cone hardening, and orbital eccentricity as
simultaneous ingredients. We identify such a joint analysis as an
important direction for future work.

\section{SMBH binary merger rate}\label{SMBHB_merger_rate}

\begin{figure*}
    \centering
    \includegraphics[width=0.49\textwidth]{A-gamma-m-21_M8.5.pdf}\hfill
    \includegraphics[width=0.49\textwidth]{A-gamma-m-2-21_M8.5.pdf}
    \includegraphics[width=0.49\textwidth]{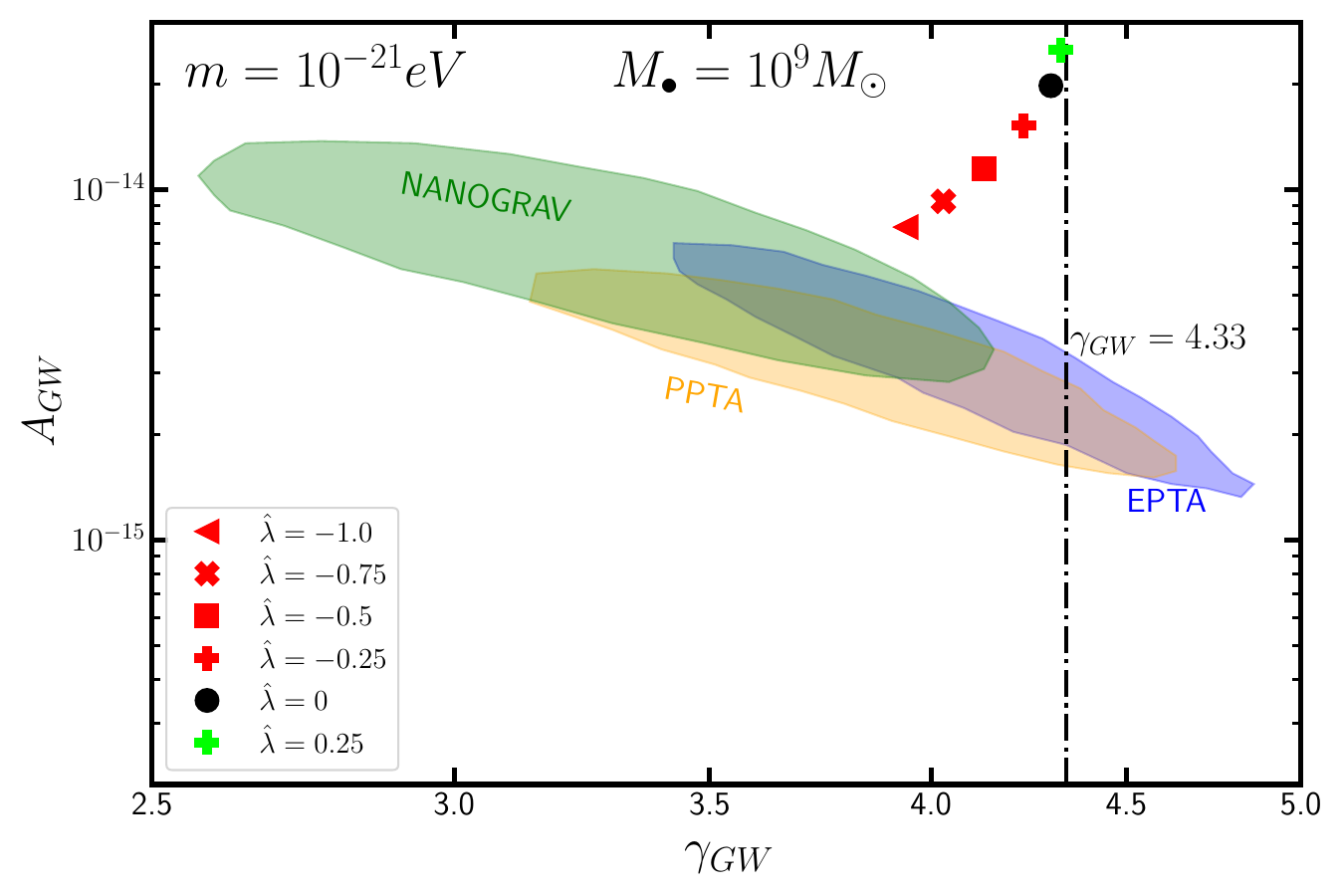}\hfill
     \includegraphics[width=0.49\textwidth]{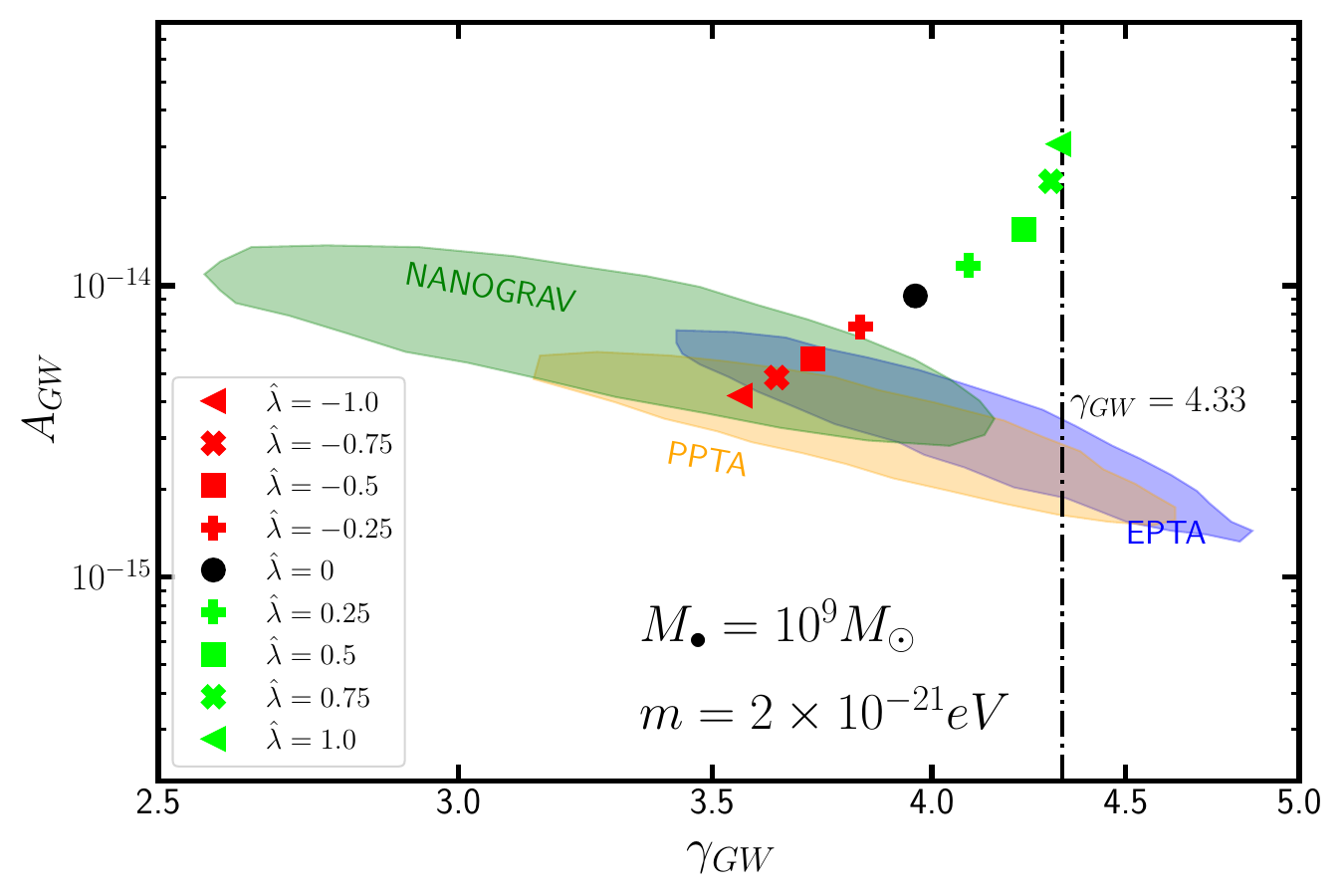}
    \caption{ Correlation between the GW amplitude (\( A_{GW} \)) and the GW slope parameter (\( \gamma_{GW} \)) for different ULDM self-coupling parameters. The shaded regions represent the correlation observed in recent PTA data \cite{DeRocco:2023qae} within 2$\sigma$ confidence level for a GW signal. The overlaid points correspond to GW strains derived from solitonic profiles with varying \( \lambda \) parameters. The top panel shows results for a equal mass binary populations where each SMBH has a mass of  \( 10^{8.5} M_{\odot} \), while the bottom panel corresponds to each SMBH having a mass of \( 10^{9} M_{\odot} \). The vertical dot-dashed line represents the value of \( \gamma_{GW} \), where the GW strain follows the power-law relation \( h_c \propto f^{-2/3} \). The reference frequency $f_{ref}$ is taken to be 1 nHz in  Eq.~\ref{GW_powerlaw} to estimate the correlated data points corresponding to different strains of solitonic density profiles.}
    \label{fig:three_plots}
\end{figure*}

Here we will shortly describe the parameters needed to calculate the number of merger events between SMBH binaries from the galaxy merger rate, available in the literature\cite{Sesana:2013wja,Sesana:2016yky,Chen:2016zyo,Chen:2018znx,NANOGrav:2024nmo}. As detailed in \cite{Chen:2018znx}, the differential galaxy merger rate per unit redshift, mass and mass ratio, can be written as,
\begin{equation}
\frac{d^{3}n_{G}}{dz dM_{G} dq}=\frac{\mathcal{\phi}(M_{G},z)}{M_{G}\log 10}\frac{\Upsilon(M_{G},z,q)}{\tau(M_{G},z,q)}\frac{dt}{dz}    
\end{equation}
where $\mathcal{\phi}(M_{G},z)$ is galaxy stellar mass function(GSMF), $\Upsilon(M_{G},z,q)$ is the differential pair function and $\tau(M_{G},z,q)$ is the merger timescale. Here $M_{G}$ is the primary galaxy mass, q is the mass ratio of galaxy pair and $z$ is the redshift of the galaxy pair. This galaxy merger rate can be converted into the SMBH binary merger rate through

\begin{equation}
\frac{dn_s}{dzdX}\equiv \frac{d^{3}n}{dzdM_{BH}dq_{BH}}=\frac{d^{3}n_{G}}{dz dM_{G} dq}\frac{dM_{G}}{dM_{BH}}\frac{dq}{dq_{BH}}
\end{equation}

The galaxy stellar masses need to be translated into SMBH masses. The total mass of a galaxy $M_{G}$ can be converted into its bulge mass $M_{bulge}$ using a phenomenological fitting function \cite{Bernardi2012SystematicEO,Sesana:2016yky},

\begin{equation}\scriptsize
\frac{M_{bulge}}{M_{G}} =
\begin{cases} 
\frac{\sqrt{6.9}}{(\log M_{G}-10)^{1.5}} 
\exp\left(\frac{-3.45}{\log M_{G}-10}\right) + 0.615, 
& \text{if } \log M_{G} > 10, \\
0.615, & \text{if } \log M_{G} \leq 10.
\end{cases}
\end{equation}

Now according to \cite{Kormendy:2013dxa}, the scaling between the galaxy bulge mass $M_{bulge}$ and the black hole mass $M_{BH}$ is of the form 
\begin{equation}
    \log_{10}\left(\frac{M_{BH}}{M_{\odot}}\right)=\mu+\alpha_{*}\log_{10}\left(\frac{M_{bulge}}{10^{11}M_{\odot}}\right)+\mathcal{N}(0,\epsilon)
\end{equation}
Here $\mathcal{N}$ is the normal distribution with mean 0 and standard deviation $\epsilon$, used to translate galaxy mass into BH mass. The black hole mass ratio $q_{BH}$ is related to galaxy mass ratio q through $q_{BH}=q^{\alpha_{*}}$.
The galaxy stellar mass function is written as 
\begin{equation}
    \mathcal{\phi}(M_{G},z)= 10^{\phi_0 +\phi_Iz}\left(\frac{M_{G}}{M_{0}}\right)^{1+\alpha_0+\alpha_Iz}\exp\left(\frac{-M_{G}}{M_{0}}\right)\ln 10
\end{equation}
The five parameters $\phi_o,\phi_I,M_0,\alpha_0,\alpha_I$ are sufficient to describe the GSMF. The differential galaxy pair function is given by,
\begin{equation}
    \Upsilon(M_{G}M_{G},z,q)=f_0' \left(\frac{M_{G}}{10^{11}M_{\odot}}\right)^{\alpha_f}(1+z)^{\beta_f}q^{\gamma_f}
\end{equation}

The quantity $f_0$ is related to $f_0'$ through $f_0=f_0'\int q^{\gamma_f}df$. So there are four model parameters $f_0', \alpha_f,\beta_f,\gamma_f$.
The merger timescale is written as, 
\begin{equation}
    \tau(M_{G},z,q)=\tau_0 \left(\frac{h_0 \times M_{G}}{0.4\times10^{11}M_{\odot}}\right)^{\alpha_\tau}(1+z)^{\beta_\tau}q^{\gamma_\tau}
\end{equation}
Here, $\tau_0,\alpha_\tau,\beta_\tau,\gamma_\tau$ are another four model parameters. The quantity $\frac{dt}{dz}$ is given assuming a flat Lambda CDM model as,
\begin{equation}
 \frac{dt}{dz}=\frac{1}{H_0 (1+z)\sqrt{\Omega_M (1+z)^{3} + \Omega_{k}(1+z)^{2}+\Omega_\Lambda}}
\end{equation}
Here, Hubble parameter is  $h_0 = 0.7$ and Hubble constant $H_0 = 70\,\, \rm km Mpc^{-1}s^{-1}$ and energy density ratios $\Omega_M = 0.3$,$\Omega_k = 0$, and $\Omega_\Lambda = 0.7$.

Summarizing, the SMBH binary merger rate is described using 18 empirical parameters that are estimated through astrophysical observations(See \cite{Chen:2018znx} for details). The SGWB signal is most sensitive for primary blackhole masses in between $10^{8}\rm M_{\odot}$ and $10^{9}\rm M_{\odot}$ and redshift parameter $z\lesssim 0.3$. The SMBH binary merger rate is typically estimated using the Holodeck library \cite{NANOGrav:2023hfp}.

\section{Plots of GW-strain amplitude vs $\gamma_{GW}$ parameter space }\label{GW_A_vs_gamma}

An alternative way to analyze the self-coupling of ULDM is through the correlation between the GW amplitude, $A_{GW}$ and the slope parameter of the GW, $\gamma_{GW}$. As discussed in \cite{DeRocco:2023qae}, recent PTA observations reveal a correlation between these parameters, represented by the shaded regions in Fig.~\ref{fig:three_plots}. Overlaid on these regions are data points derived from GW strains corresponding to solitonic profiles with different $\lambda$ values. The figures indicate that self-coupling is most effectively probed for a ULDM mass of $\sim 10^{-21}$eV, whereas for lighter masses, the signal shifts beyond the correlated regions detectable by current PTA observations. Additionally, for an SMBH of mass $10^{9}\rm M_{\odot}$, the strain amplitude increases compared to that for an SMBH of $10^{8.5}\rm M_{\odot}$ ,  consistent with merger calculations. The vertical dot-dashed line represents the value of  $\gamma_{GW}$ in the absence of a DM environment around the SMBH merger, where the GW strain follows the expected power-law behaviour $h_{c}\propto f^{-2/3}$.

\bibliographystyle{JHEP}       
\bibliography{ref-soliton}     
\end{document}